\documentclass[10pt,twocolumn,twoside]{IEEEtran}
\def\ps@headings{%
\def\@oddhead{\mbox{}\scriptsize\rightmark \hfil \thepage}%
\def\@evenhead{\scriptsize\thepage \hfil \leftmark\mbox{}}%
\def\@oddfoot{}%
\def\@evenfoot{}}
\makeatother 

\usepackage{amsmath,amssymb,mathrsfs,amsthm,bm}
\usepackage{mathtools}
\usepackage{cite}
\usepackage{algorithm}
\usepackage{tikz}
\usetikzlibrary{arrows.meta,positioning,calc}
\usepackage{pgfplots}
\usepgfplotslibrary{groupplots,fillbetween}
\pgfplotsset{compat=1.18}

\definecolor{rodpblue}{rgb}{0.1,0.2,0.6}
\definecolor{tdpred}{rgb}{0.8,0.2,0.1}
\definecolor{zgreen}{rgb}{0.1,0.5,0.1}
\definecolor{totalblue}{rgb}{0.3,0.4,0.8}

\definecolor{zoneIM}{rgb}{0.95,0.95,0.98}
\definecolor{zoneNZ}{rgb}{0.92,0.96,0.92}
\definecolor{zoneEX}{rgb}{0.98,0.94,0.94}
\usepackage{lipsum}
\usepackage{graphicx,stfloats,subfigure,multicol}%amsmath
\usepackage{epsfig,epsf,psfrag,amssymb,amsfonts,latexsym,graphicx,mathrsfs,subfigure}
\usepackage[table,xcdraw]{xcolor}
\usepackage{lettrine}
\usepackage{comment}
\usepackage{enumitem}
\newlist{enumsteps}{enumerate}{2}
\setlist[enumsteps,1]{label=Case \arabic*: }
\setlist[enumsteps,2]{label=Case \arabic{enumstepsi}.\arabic*: }
\usepackage{booktabs}
\usepackage[table]{xcolor}
\usepackage{multirow}
\usepackage{tablefootnote}
\usepackage{bbm}
\usepackage{orcidlink}
\usepackage{chngpage}
\usepackage{textcomp}
\usepackage{hyperref}
\usepackage{pifont}
\usepackage{url}
\usepackage[hyphenbreaks]{breakurl}
\usepackage[normalem]{ulem}
\usepackage{algpseudocode}
\newsavebox{\ieeealgbox}

\usepackage{array}

\newtheorem{theorem}{Theorem}
\newtheorem{proposition}{Proposition}
\newtheorem{corollary}{Corollary}
\newtheorem{lemma}{Lemma}

\newtheorem*{policy*}{Dynamic NEM}
\newtheorem*{policy1*}{Generalized Dynamic NEM}

 \def\old#1{}

\def\beq{\begin{equation}}
\def\eeq{\end{equation}}
\def\bea{\begin{eqnarray}}
\def\eea{\end{eqnarray}}
\def\ba{\begin{array}}
\def\ea{\end{array}}

\def\bitem{\begin{itemize}}
\def\eitem{\end{itemize}}
\def\ben{\begin{enumerate}}
\def\een{\end{enumerate}}

\def\ie{{\it i.e.,\ \/}}

\definecolor{bgrd}{rgb}{1,1,1}
\definecolor{gray}{rgb}{0.5,0.5,0.5}
\definecolor{dkr}{rgb}{0.7,0.1,0.2}
\definecolor{dkb}{rgb}{0.1,0.1,0.8}

\def\tcb{\textcolor{blue}}

\newcommand{\mbbE}{\mathbb{E}}

\newcommand{\mbbR}{\mathbb{R}}

\def\Bc{{\cal B}}

\def\Lc{{\cal L}}

\def\Pc{{\cal P}}
\def\Qc{{\cal Q}}

\def\Tc{{\cal T}}

\def\Wc{{\cal W}}

\DeclareMathOperator{\sgn}{sgn}

\usepackage{microtype}

\begin{document}

\title{Optimal Scheduling of Electricity and Water in Renewable-Colocated Desalination Plants
}

\author{Ahmed~S.~Alahmed\orcidlink{0000-0002-4715-4379},
Audun~Botterud\orcidlink{0000-0002-5706-5636}, and
Saurabh~Amin\orcidlink{0000-0003-1554-015X}
\thanks{\scriptsize  Ahmed S. Alahmed ({\tt {\tcb{alahmad@kfupm.edu.sa}}}) is with King Fahd University of Petroleum and Minerals, Dhahran, KSA. Audun Botterud and Saurabh Amin ({\tt {\tcb{audunb,amins}\}\tcb{@mit.edu}}}) are with Massachusetts Institute of Technology, Cambridge, MA.}
} 

\maketitle
\begin{abstract}
We develop a mathematical framework for the optimal scheduling of flexible water desalination plants (WDPs) as hybrid generator-load resources. WDPs integrate thermal generation, membrane-based controllable loads, and renewable energy sources, offering unique operational flexibility for power system operations. They can simultaneously participate in two markets: selling desalinated water to a water utility, and bidirectionally transacting electricity with the grid based on their net electricity demand. We formulate the scheduling decision problem of a profit-maximizing WDP, capturing operational, technological, and market-based coupling between water and electricity flows. The threshold-based structure we derive provides computationally tractable coordination suitable for large-scale deployment, offering operational and economical insights into how thermal and membrane-based desalination colocated with renewables complementarily provide continuous bidirectional flexibility. The thresholds are analytically characterized in near closed form as explicit functions of technology and tariff parameters. We examine how small changes in the exogenous tariff and technology parameters affect the WDP's profit. Extensive simulations illustrate the optimal WDP's operation, profit, and water-electricity exchange, demonstrating significant improvements relative to benchmark algorithms.
\end{abstract}


\vspace{-0.3cm}
\section{Introduction}\label{sec:intro}
\lettrine{T}{he} ongoing transformation of power systems, driven by increasing penetration of variable renewables, creates growing demand for flexible resources that can provide bidirectional grid services. Water desalination plants (WDPs),\footnote{By WDP, we generally also refer to water treatment plants, which are typically modeled similarly but with different salinity-related parameters \cite{Oikonomou&Parvania:20TSG}.} representing approximately 5-12\% of total electricity consumption in Middle East countries \cite{Siddiqi&Anadon:11AE} with 7\% per annum global capacity growth since 2010 \cite{Eke&Yusuf&Giwa&Sodiq:20Desalination}, are evolving from passive to active market participants capable of dynamically adjusting operations in response to price signals and renewable energy availability. Unlike conventional generators that can only supply power or loads that can only reduce consumption, modern WDPs integrating both thermal and renewable generation and membrane-based loads function as hybrid generator-load resources with continuous bidirectional flexibility.

With global freshwater demand projected to exceed available supply by 40\% by 2030 \cite{TurningTheTide:23GlobalWaterCommission}, WDPs have become an increasingly vital component of sustainable water infrastructure in various parts of the world. As the demand for seawater and wastewater treatment continues to grow, so does the need for efficient and sustainable operation of WDPs. These facilities are inherently energy-intensive, consuming substantial amounts of energy to drive both thermal- and membrane-based desalination processes \cite{Elimelech:2011Science}. Consequently, there is growing interest in integrating desalination operations with renewable energy sources and advanced scheduling strategies to reduce operational costs and environmental impact \cite{Ghaffour:2015Desalination}.

In parallel, the ongoing transformation of power systems, driven by increasing penetration of renewables and the rise of decentralized energy resources, has given rise to novel opportunities and challenges at the water-energy nexus. In this context, WDPs are evolving from passive energy consumers to active participants in electricity markets, capable of dynamically adjusting their operations in response to electricity price signals and renewable energy availability. Co-optimizing water and power scheduling, a global research priority in the water-energy nexus \cite{ONeiletal:24OAJPE}, is crucial for ensuring economic viability and grid-supportive behavior.

\begin{figure}[t]
    \centering
    \includegraphics[scale=0.70]{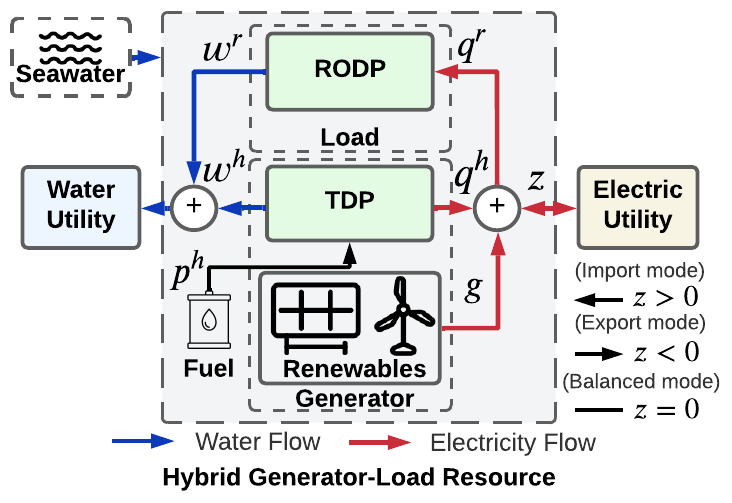}
    \vspace{-0.4cm}
    \caption{WDP as a hybrid generator-load resource. The water (power) output of the thermal desalination plant (TDP) and reverse osmosis desalination plant (RODP) is denoted by $w^h, w^r\; (q^h, q^r) \in \mbbR_+$, respectively. The fuel consumption of TDP is denoted by $p^h \in \mbbR_+$. The renewable generation and net consumption are denoted by $g \in \mbbR_+$, and $z \in \mbbR $.}
    \label{fig:WDP}
\end{figure}

This paper develops an analytical framework for the optimal scheduling of WDPs operating as hybrid generator-load resources (Figure \ref{fig:WDP}). A key challenge lies in utilizing the complementarities between the load-generator resources within the plant, in addition to optimally scheduling both water and electricity transactions with water and electric utilities.

\vspace{-0.35cm}
\subsection{Related Work}
One of the earliest examinations of water-power systems is Gleick’s seminal review \cite{Gleick:94ARRR}, which highlighted the fundamental interdependence between water and power systems: energy is required for water extraction, treatment, and transport, while water is essential for electricity generation and fuel production. This foundational perspective has since motivated a broad literature on modeling, coordinating, and optimizing coupled water-energy infrastructures. 

In energy-abundant regions, TDPs often operate alongside power generation, producing freshwater as the primary output and electricity as a by-product. These colocated systems leverage shared thermal and electrical infrastructure, resulting in strongly coupled operational behavior. For instance, \cite{Santhosh&Farid&Toumi:14AE} propose dynamic, physics-based models of such integrated systems, emphasizing the value of coordinated control between thermal power and thermal desalination units.

Reverse osmosis (RO) has emerged as the dominant desalination technology, favored for its modularity and superior energy efficiency. Recent estimates suggest that RO accounts for approximately 69\% of global installed desalination capacity \cite{Eke&Yusuf&Giwa&Sodiq:20Desalination}. As RO units rely predominantly on electricity from the grid, they can function as substantial and flexible loads within the power system. Several recent studies \cite{Ghaithan2022,Qudah2024,Almehizia&Almasri&Hussein&Ehsani:19TSG,Guoetal:16TSG} explore the integration of RODPs with variable renewable energy sources (e.g., solar and wind), primarily through simulation-based energy management frameworks designed to reduce operational costs, emissions, and improve grid compatibility. In particular, \cite{Guoetal:16TSG} present an energy management system for a standalone wind-powered desalination microgrid, highlighting coordinated control strategies for matching variable generation to water demand. However, these contributions remain largely simulation-based and typically treat the power system as an exogenous input, without explicitly modeling the endogenous coupling between the water and energy sectors. More recent efforts \cite{MoazeniAPEN2020,Moazeni&Khazaei&Asrari:21TSG,AlAwamiTPS2022} address this gap by developing numerical co-optimization models for water-energy microgrids and joint RO-power dispatching under dynamic demand, grid constraints. These studies also explore the role of RO in demand response and grid services. The work in \cite{Mohammadi&Ardakani&Abdullah&Heydt&Thomas:19TPS} formulates a security-constrained unit commitment problem to coordinate the operation of RO-only desalination plants and power plants and showed a reduction in the system's operating costs by 5\% to 8\% under the IEEE 118-bus system.

Despite these advances, existing approaches are predominantly computational, with limited analytical insight into structural properties of the optimal scheduling problem, an issue that our work aims to address.

Beyond RO-power coordination, recent work expands to decentralized water-energy nexus markets and flexible cross-sector operations. \cite{ZhaoetalBlockchain:23TSG} propose a blockchain-based transactive framework for interconnected water-energy hubs, showing performance gains under spatial-temporal uncertainty. Similarly, \cite{CaoetalWatershed:23TSG} develops a watershed-electricity model that exploits hydrological flexibility, through dynamic river conditions and pump scheduling, to support distribution network operations. While demonstrating the benefits of decentralized coordination, these studies do not consider colocating and the unique operating regimes of joint RO and TDP systems.

The studies in \cite{Oikonomou&Parvania:20TSG, Zamzam&Emiliano&Zhao&Taylor&Sidiropoulos:19TCNS,Gu&Sioshansi:25PESGM,Li&Yu&Sumaiti&Turitsyn:19TCNS} propose models for coordinating water and power distribution systems to minimize operational costs of meeting water and electricity demands by leveraging variable speed pumps and water tanks. However, these models do not incorporate colocated renewable energy sources or bidirectional energy transactions between WDPs and the electric grid, limiting their ability to fully optimize integrated resource management.

Although significant progress has been made, most existing work models RODP and TDP separately, without capturing their distinct operational features or interactions with water and power systems, particularly when colocated with renewables. This siloed treatment overlooks opportunities for coordinated real-time scheduling and efficient resource utilization. In addition, existing studies are predominantly numerical- and simulation-based, whereas we develop an analytical framework that enables scalable solutions with structural insights, including operational intuitions and comparative statics.

\subsection{Main Contributions}
Our main contribution is the analytical characterization of the optimal electricity-water scheduling for WDPs operating as hybrid generator-load resources, enabling efficient grid integration while maintaining water production requirements. The threshold-based structure we derive provides computationally tractable coordination suitable for large-scale deployment, offering operational insights into how thermal generation and membrane-based loads complementarily provide continuous bidirectional flexibility.

From a power grid operations perspective, our results demonstrate how WDPs can provide: (i) bidirectional dispatch flexibility (import, balanced (net-zero), export modes) without energy storage; (ii) combined response from coordinated generation-load control; (iii) optimal dispatch through thresholds computable offline and independent of renewables; (iv) smooth operational transitions at constant zero exchange with the electricity grid; and (v) a real-time receding-horizon implementation that accommodates renewable uncertainty while preserving the threshold structure.

\begin{figure}[t]
    \centering
    \includegraphics[width=\linewidth]{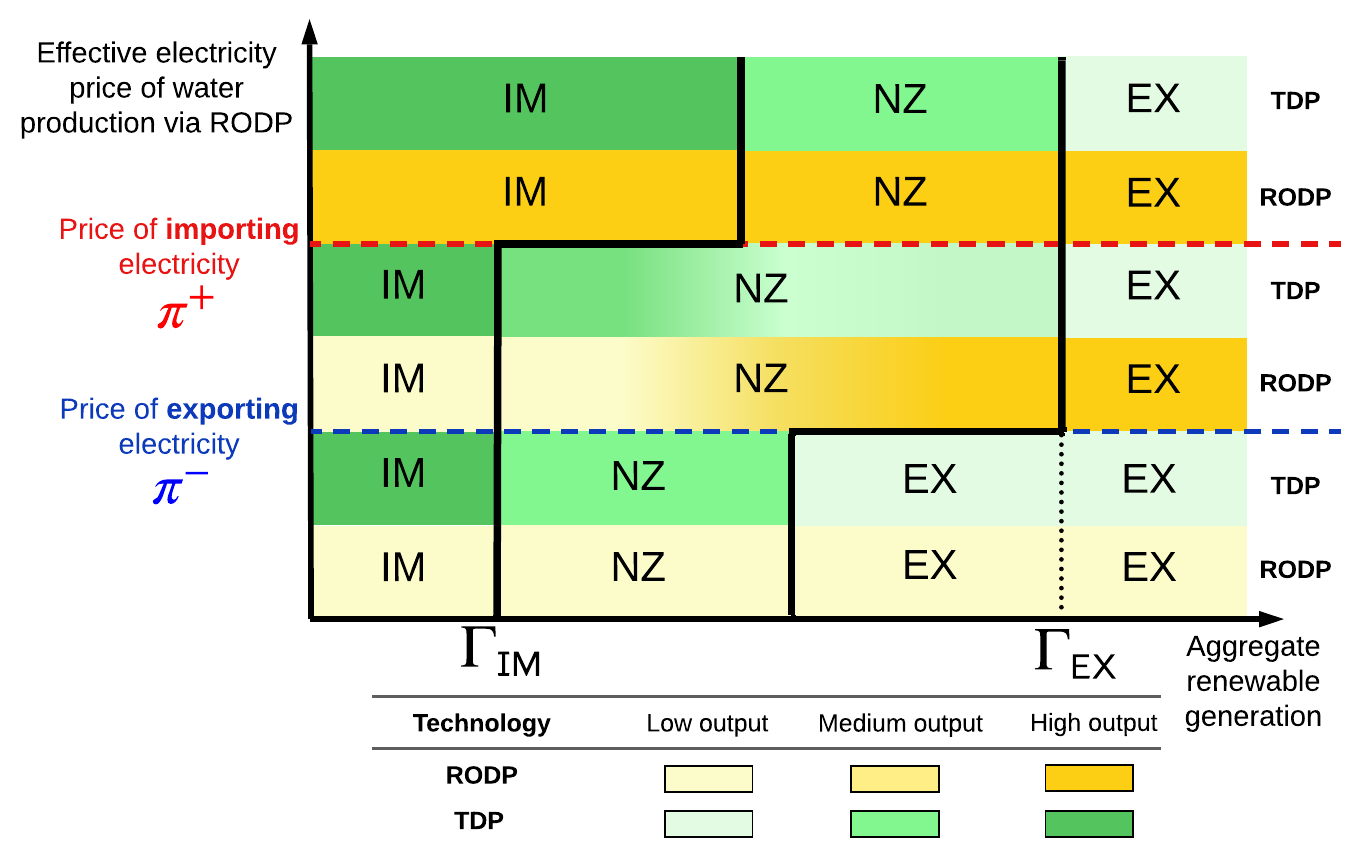}
    \vspace{-0.8cm}
    \caption{Optimal WDP water dispatch across electricity operating modes (IM: import, NZ: net-zero, EX: export).}
    \label{fig:ProductionPlan}
\end{figure}

Figure \ref{fig:ProductionPlan} shows the WDP's optimal water dispatch, also determining power dispatch, based on aggregate renewable generation, effective electricity price of RODP water, and grid prices of imports ($\pi^+$) and exports ($\pi^-$), where $\pi^+ \geq \pi^-$. The WDP’s optimal operation is segmented into three modes (IM, NZ, EX) depending on renewables' output by two thresholds ($\Gamma_{\sf{IM}},\Gamma_{\sf{EX}}$). It operates as a controllable load in the (IM) mode, a net-zero self-sufficient microgrid in the (NZ) mode, and a dispatchable generator in the (EX) mode.

As illustrated in Figure \ref{fig:ProductionPlan}, the TDP’s optimal water and power outputs decrease monotonically with the availability of local renewables, whereas the RODP’s water output increases monotonically with renewables. If the effective electricity price of water production via RODP  exceeds $\pi^+$, the RODP operates at its maximum capacity, as the marginal revenue from water sales exceeds the marginal cost of power, thus expanding the mode (IM). In contrast, if the effective electricity price of water production via RODP is less than $\pi^-$, the RODP is scheduled to reach its minimum capacity, expanding the (EX) mode. In both boundary cases, TDP output adjusts to renewable availability according to a two-threshold policy. In the net-zero power-balanced (NZ) mode, TDP generation follows renewable-adjusted RODP consumption.

The most pertinent scenario arises when the effective electricity price of RODP water production exceeds the export price $\pi^-$ but remains below the import price $\pi^+$. Under this condition, we demonstrate that the optimal schedules for TDP and RODP are coupled and also abide by a threshold policy. As onsite renewables increase, both TDP and RODP dynamically adjust to absorb surplus generation, expanding the power-balanced (NZ) mode while shrinking the power-importing (IM) and exporting (EX) modes.

In addition to the general structure of the solution, we analyze the optimal WDP dispatch in cases of single desalination technology. Also, we show the effect of the exogenous water and electricity tariff and technology parameters on the WDP's maximum profit. We provide extended simulation results illustrating the optimal WDP’s profit, scheduling, and patterns of water and electricity exchange compared to benchmarks.

To handle renewable uncertainty in real-time operation, we develop a model-predictive dispatch algorithm whose per-step decision reduces to an instance of the offline threshold policy in Theorem~\ref{thm:optimal}. This yields a tractable scheme that avoids the curse of dimensionality inherent in the dynamic-programming solution of the underlying multistage stochastic problem, with suboptimality governed by the renewable forecast error.

\vspace{-0.4cm}
\subsection{Major Mathematical Symbols and Paper Organization}
% For vector case:

%Table \ref{tab:MajorSymbols} lists the major symbols used. The use of notations here is standard. When necessary, boldface letters denote column vectors as in $\bm{x}=(x^1,\cdots, x^n)$ and $\bm{x}^\top$ its transpose. Let $\bm{1}_n \in \mbbR^n$ denote the length-$n$ vector of all ones. For vectors $\bm{x},\bm{y}$,  $\bm{x} \preceq \bm{y}$ is the element-wise inequality $x^i \le y^i$ for all $i$, and $[\bm{x}]^+, [\bm{x}]^-$ are the element-wise positive and negative parts of vector $\bm{x}$, \ie $[x^i]^+=\max\{0,x^i\}$, $[x^i]^- =-\min\{0,x^i\}$ for all $i$, and $\bm{x}= [\bm{x}]^+ - [\bm{x}]^-$. Max and min operators applied to column vectors (e.g., $\max\{\bm{0}, \bm{x}\}$) are taken elementwise; that is, $\max\{\bm{0}, \bm{x}\} = (\max\{0, x^1\}, \ldots, \max\{0, x^n\})$. Lastly, we use $\mbbR_+$ to denote the set of non-negative real numbers, \ie $\mbbR_+=\{ x\in \mbbR: x\geq 0\}$. 

% For scalar case:
Table \ref{tab:MajorSymbolsGrouped} lists the major symbols and notations used. The WDP system model and scheduling problem are presented in $\S$\ref{sec:form}, followed by a delineation of the optimal scheduling and its intuitions in $\S$\ref{sec:Optimal}. Comparative statics of the optimal scheduling policy is provided in $\S$\ref{sec:SpecialCase}. Detailed simulations are shown in $\S$\ref{sec:num}, followed by the conclusion and future directions in $\S$\ref{sec:conclusion}. Appendix~\ref{app:MPC} presents a receding-horizon algorithm that handles renewable uncertainty in real-time operation.

\begin{table}[ht]
\centering
\caption{WDP Major variables and parameters}
\label{tab:MajorSymbolsGrouped}
\vspace{-0.2cm}
\resizebox{\columnwidth}{!}{%
\begin{tabular}{@{}ll@{}}
\midrule \midrule
\textbf{Symbol} & \textbf{Description} \\ \midrule
\multicolumn{2}{@{}l}{\textbf{Sets and Indices}} \\
$\Bc$ & Billing period. \\
$\Tc, t$ & Dispatch horizon and decision period $t \in \Tc$. \\[0.3em]
\multicolumn{2}{@{}l}{\textbf{Decision and Exogenous Variables}} \\
$g_t$ & Renewable generation output at period $t$. \\
$s$ & Aggregate renewable generation over $\Bc$. \\
$q_t^h, q_t^r$ & Electricity prod. (consumption) of TDP (RODP). \\
$w_t^h, w_t^r$ & Water output of TDP and RODP. \\
$p_t^h$ & TDP fuel consumption. \\
$z$ & Net electricity exchange. \\[0.3em]
\multicolumn{2}{@{}l}{\textbf{Technical Parameters and Functions}} \\ 
$\eta^h$ & TDP water-to-electricity production ratio. \\
$f^h(\cdot), f^r(\cdot)$ & Conversion functions of TDP and RODP. \\
$\phi^h(\cdot), \phi^r(\cdot)$ & Inverse of $f_t^{\prime h}$ and $f_t^{\prime r}$. \\
$\mathcal{Q}(\cdot)$ & Aggregate controllable net consumption. \\
$\widetilde{w}^h(\cdot)$ & TDP stationarity function. \\
$\underline{w}^h, \overline{w}^h$ & Minimum and maximum water flowrates of TDP. \\
$\underline{w}^r, \overline{w}^r$ & Minimum and maximum water flowrates of RODP. \\
$\underline{q}^r, \overline{q}^r$ & Minimum and maximum elec. input of RODP. \\
$\Gamma_{\sf{IM}}, \Gamma_{\sf{EX}}$ & Mode-defining optimal scheduling thresholds. \\
$\Gamma_{\sf{NZ_1}}, \Gamma_{\sf{NZ_2}}$ & Net-zero mode optimal scheduling thresholds. \\[0.3em]
\multicolumn{2}{@{}l}{\textbf{Prices, Costs, and Profit}} \\ 
$C^h(\cdot), (C^h)'(\cdot)$ & TDP cost function and its derivative. \\
$D^h(\cdot)$ & Inverse of the TDP marginal cost $(C^h)'(\cdot)$. \\
$\Pi(\cdot)$ & WDP profit function. \\
$P(\cdot), R^{w}(\cdot)$ & Electricity (Water) payment (revenue) function. \\
$\pi^+, \pi^-$ & Electricity retail (import) and sell (export) prices. \\
$\pi^w$ & Water selling price. \\
$\mu$ & Dual variable of the energy balance. \\
$\delta_{\sf{NZ_1}}, \delta_{\sf{NZ_2}}$ & Effective RODP water prices at $\underline{q}^r, \overline{q}^r$. \\[0.3em]
\multicolumn{2}{@{}l}{\textbf{Mathematical Notations}} \\
$\mbbR_+, \mbbR_{++}$ & Sets of nonnegative and positive real numbers.\\
$[x]^+,[x]^-$ & $\max\{0, x\}$ and $-\min\{0, x\}$, for any \( x \in \mbbR \)\\
$[x]_{[a,b]}$ & $\max\{a, \min\{x, b\}\}$, with \(x, a, b \in \mbbR \) and \( a \leq b \) \\
\midrule \midrule
\end{tabular}%
}
\end{table}

\section{WDP Framework and Dispatch Problem}\label{sec:form}
From a grid operator perspective, the WDP represents a controllable  
resource capable of both generation and consumption. Unlike conventional 
resources with unidirectional power flow, the WDP can smoothly transition 
between grid import, balanced, and export states through coordinated 
control of TDP and RODP units. This section formalizes the WDP's 
operational characteristics and scheduling problem.

We consider the problem of a WDP that jointly optimizes water and 
electricity schedules to maximize its profits from transacting the two 
commodities with water and electric utilities (Fig.~\ref{fig:WDP}). The 
WDP integrates thermal and membrane-based desalination units, colocated 
with renewables on-site. It engages in a unilateral transaction with the 
water utility, supplying water at an exogenously specified 
price, which may vary over time. In contrast, its interaction with the 
electric utility is bidirectional: the WDP imports electricity when 
its demand exceeds local generation and exports surplus power otherwise. 
This arrangement is analogous to virtual power plants 
\cite{Koraki&Strunz:18TPS} and grid-connected microgrids 
\cite{Jiang&Xue&Geng:13TPS}. In both markets, 
the WDP is modeled as a price-taking agent.

As shown in Figures~\ref{fig:WDP} and~\ref{fig:ProductionPlan}, the WDP 
exhibits a noteworthy complementarity between water and electricity 
systems. Specifically, the TDP uses a fuel source to co-generate both 
water and electricity, while the RODP consumes electricity to produce 
water. As a result, the electrical load within the WDP is the RODP, 
whose demand is met through a combination of local generation from the 
TDP, on-site renewables, and electricity imported from the grid, with 
contributions varying based on availability and operating conditions.

The WDP's operational timescale is determined by its measurement and 
scheduling frequency, whereas transactions with the water and electric 
utilities are governed by their billing periods (e.g., 15 minutes or 
one hour). At the beginning of each billing period $\Bc$, the water and 
electricity prices are announced to the WDP and remain fixed throughout 
$\Bc$. Given these prices, the WDP schedules its resources over the 
scheduling horizon $\Tc := \{1, \ldots, T\} \subseteq \Bc$.

Next, we present the modeling of the TDP, RODP, payment functions to the 
water and electricity utilities, and the WDP scheduling decision problem.

\vspace{-0.25cm}
\subsection{WDP Resources}

\subsubsection{TDP}
TDPs use a boiler to heat seawater using a fuel source,\footnote{Fuel 
sources may include fossil-based (e.g., natural gas, oil), nuclear, 
hydrogen-based, or other energy sources.} producing steam that is then 
condensed to obtain fresh water. The waste heat is sourced to produce electricity. TDPs, therefore, produce both water and electricity.

At every decision period $t \in \Tc$, the TDP consumes fuel, denoted by  $p_t^h$, and produces both water and electricity, which we denote by  $w_t^h \in \mbbR_+$ and $q_t^h \in \mbbR_+$, respectively. The TDP desalinated water is related to fuel consumption via 
$w_t^h = f_t^h(p_t^h)$, where $f_t^h: \mbbR_+ \to \mbbR_+$ is strictly concave, continuously differentiable, and strictly increasing. We denote the marginal of $f_t^h$ at $p_t^h$ by $f_t^{\prime h}(p_t^h) := \partial f_t^h(p_t^h) / \partial p_t^h$, and its inverse by $\phi_t^h := (f_t^{\prime h})^{-1}$. We require 
$f_t^{\prime h} > 0$ throughout the operating range, so that $\phi_t^h$ is well defined and differentiable there.

For every $t \in \Tc$, the TDP co-produces water and electricity according 
to the relation $w_t^h = \eta_t^h q_t^h$, where $\eta_t^h \in \mbbR_{++}$ 
is the water-to-electricity production ratio ~\cite{Santhosh&Farid&Toumi:14AE}. Define 
$\beta_t^h(p_t^h) := f_t^{\prime h}(p_t^h) / \eta_t^h$, the marginal 
conversion factor from fuel to electricity. The TDP 
water flow rate is bounded by minimum and maximum operating limits 
$\underline{w}^h, \overline{w}^h \in \mbbR_+$ that ensure stable 
operation, \ie $w_t^h \in [\underline{w}^h, \overline{w}^h]$.

For each $t \in \Tc$, the TDP's operating cost function is denoted by 
$C_t^h(p_t^h) \in \mbbR_+$, which depends on the fuel cost and plant 
efficiency. We assume $C_t^h(\cdot)$ to be strictly convex, continuously 
differentiable, and non-decreasing \cite{Santhosh&Farid&Toumi:14AE}, standard for thermal fuel costs \cite{Wood&Wollenberg&Sheble:13Wiley}. Let $D_t^h(\cdot)$ denote the inverse 
of $\partial C_t^h(p_t^h) / \partial p_t^h$, which is well-defined by 
strict convexity. 

\subsubsection{RODP}
The RODP consumes electricity to force pressurized seawater (or 
wastewater) molecules through a semi-permeable membrane under high 
pressure, effectively filtering out salt and 
impurities~\cite{Mulder:96Springer}.

For every $t \in \Tc$, the water production and electricity consumption of the RODP,  $w_t^r, q_t^r \in \mbbR_+$, are related as $w_t^r = f^r(q_t^r)$, where $f^r: \mbbR_+ \to \mbbR_+$ is strictly concave, continuously differentiable, and strictly increasing.\footnote{The strict concavity captures diminishing returns from over-pumping due to membrane fouling and pressure-flow nonlinearity \cite{Zhu&Christofides&Cohen:09ACSAECR}.} We denote the marginal of $f^r$ at $q_t^r$ by 
$f^{\prime r}(q_t^r) := \partial f^r(q_t^r) / \partial q_t^r$, which is 
strictly positive and strictly decreasing on the operating range, and its inverse by $\phi^r := (f^{\prime r})^{-1}$. The RODP 
water production is bounded by minimum and maximum operating limits 
$\underline{w}^r, \overline{w}^r \in \mbbR_+$, \ie 
$w_t^r \in [\underline{w}^r, \overline{w}^r]$, equivalently 
$q_t^r \in [\underline{q}^r, \overline{q}^r]$ where 
$\underline{q}^r := (f^r)^{-1}(\underline{w}^r)$ and 
$\overline{q}^r := (f^r)^{-1}(\overline{w}^r)$.

\subsubsection{Renewable Energy Sources}
The plant has colocated renewables whose generation, for each 
$t \in \Tc$, is denoted by $g_t \in \mbbR_+$. We first characterize the 
optimal dispatch when $\bm{g}$ is known (Theorem~\ref{thm:optimal}). In 
real-time operation, the components $g_t$ are revealed sequentially over 
the billing period, and each setpoint must be committed before the 
billing-period aggregate $s:=\bm{1}^\top \bm{g}$ is realized. The resulting 
uncertainty is addressed in Appendix~\ref{app:MPC} via a receding-horizon 
implementation that leverages the optimal dispatch structure and does not require the underlying probability distribution of $\bm{g}$.

Capital costs of renewable installations are excluded from the analysis, 
as we focus on short-run operation rather than long-term capacity planning.

\vspace{-0.2cm}
\subsection{WDP Water and Electricity Payments}

\subsubsection{Water Revenue}
The profit-maximizing WDP sells the total TDP and RODP desalinated water 
to the water utility. The WDP's revenue over the scheduling horizon is 
therefore
\begin{equation}\label{eq:WaterPayment}
    R^{w}(\bm{w}^h, \bm{w}^r) 
    = \pi^w \bm{1}^\top (\bm{w}^h + \bm{w}^r),
\end{equation}
where $\pi^w \in \mbbR_+$ is the water selling price (announced at the 
start of $\Bc$), and $\bm{w}^h := (w_t^h)_{t \in \Tc}$, 
$\bm{w}^r := (w_t^r)_{t \in \Tc}$ are the vectors of the TDP and RODP water outputs 
over $\Tc$.

\subsubsection{Electricity Payment}
The profit-maximizing WDP bidirectionally transacts with the electric 
utility, which charges the WDP based on its net electricity exchange 
over the billing period. The net exchange 
$z: \mbbR_+^T \times \mbbR_+^T \rightarrow \mbbR$ of the WDP is
\begin{equation}\label{eq:NetConsumption}
    z(\bm{q}^h, \bm{q}^r; \bm{g}) 
    = \bm{1}^\top (\bm{q}^r - \bm{q}^h - \bm{g}),
\end{equation}
with $\bm{q}^h := (q_t^h)_{t \in \Tc}$, $\bm{q}^r := (q_t^r)_{t \in \Tc}$, and $\bm{g} := (g_t)_{t \in \Tc}$.

The electric utility charges the WDP as an industrial customer under a 
{\em net energy metering tariff}, as
\begin{equation}\label{eq:ElectricPayment}
    P(\bm{q}^h, \bm{q}^r; \bm{g}) 
    = \pi^{+} [z(\bm{q}^h, \bm{q}^r; \bm{g})]^+ 
    - \pi^{-} [z(\bm{q}^h, \bm{q}^r; \bm{g})]^- 
    + \pi^{0},
\end{equation}
where $\pi^{+}, \pi^{-} \in \mbbR_+$ are the import and export prices, 
respectively, $\pi^{0} \in \mbbR$ is the non-volumetric fixed charge. To avoid risk-free 
arbitrage by the WDP, we assume $\pi^{+} \geq \pi^{-}$. A special case of 
\eqref{eq:ElectricPayment} arises when the WDP faces a single price 
($\pi^{+} = \pi^{-}$), e.g., a WDP exposed to locational marginal prices in competitive electricity markets. From \eqref{eq:ElectricPayment}, the WDP is {\em importing power} (IM) if 
$z > 0$, {\em exporting power} (EX) if $z < 0$, and {\em power-balanced} 
(NZ) if $z = 0$.

While we focus on a single billing period $\Bc$ with fixed prices, the 
formulation extends naturally across multiple billing periods with time-varying prices, e.g., time-of-use or real-time 
pricing~\cite{Alahmed&Tong:22IEEETSG}. Alternatively, utilities may 
establish long-term contracts with WDPs that fix import/export 
prices.

\vspace{-0.2cm}
\subsection{WDP Profit}
The WDP's profit over the billing period is
\begin{align}\label{eq:WDPprofit}
    \Pi(\bm{w}^h, \bm{w}^r, \bm{q}^h, \bm{q}^r, \bm{p}^h; \bm{g}) 
    :={}& R^{w}(\bm{w}^h, \bm{w}^r) - P(\bm{q}^h, \bm{q}^r; \bm{g}) \notag \\
    & - \sum_{t \in \Tc} C_t^h(p_t^h),
\end{align}
where the first term represents the revenue from selling water; the second term represents the monetary transaction from net electricity exchange, which constitutes a cost when electricity is imported and 
revenue when it is exported; and the third term reflects the operating 
costs of the TDP. In contrast, RODP operating costs are implicitly 
accounted for through the cost of procuring electricity $q_t^r$, which 
can be sourced from colocated renewables, the TDP, the grid, or any 
combination of them.

\vspace{-0.2cm}
\subsection{Profit-Maximizing WDP Dispatch}
The WDP jointly schedules thermal and RO units to maximize profit subject 
to production, capacity, and demand constraints; hence, it solves
\begin{align}\label{eq:Optimization}
    (\bm{w}^{h\ast}, \bm{w}^{r\ast}, \bm{q}^{h\ast}, \bm{q}^{r\ast}, \bm{p}^{h\ast}) 
    &:={}\hspace{-0.8cm} \underset{\bm{w}^h, \bm{w}^r, \bm{q}^h, \bm{q}^r, \bm{p}^h \in \mbbR_+^T}
    {\operatorname{argmax}} \hspace{-0.6cm}
    \Pi(\bm{w}^h, \bm{w}^r, \bm{q}^h, \bm{q}^r, \bm{p}^h; \bm{g}) \notag \\
    & \text{subject to} \;\;  \underline{w}^{h} \bm{1} \preceq \bm{w}^h \preceq \overline{w}^{h} \bm{1}, \notag \\
    & \hphantom{\text{subject to}} \;\; 
    \underline{w}^{r} \bm{1} \preceq \bm{w}^r \preceq \overline{w}^{r} \bm{1}, \notag \\ 
    & \hphantom{\text{subject to}} \;\; \bm{1}^\top (\bm{w}^h + \bm{w}^r) \geq \Wc, \notag \\
    & \hphantom{\text{subject to}} \;\; w_t^r = f^r(q_t^r), \;\; \forall t \in \Tc, \notag \\
    & \hphantom{\text{subject to}} \;\; w_t^h = \eta_t^h q_t^h,  \;\; \forall t \in \Tc, \notag \\
    & \hphantom{\text{subject to}} \;\; w_t^h = f_t^h(p_t^h), \;\; \forall t \in \Tc, 
\end{align}
where $\Wc \in \mbbR_+$ is the water demand over the billing 
period. We assume the WDP is sized so that the minimum output meets 
demand, \ie $T(\underline{w}^h + \underline{w}^r) \geq \Wc$. The problem 
in~\eqref{eq:Optimization} is well-posed with a global optimum, as the 
concave objective (Lemma~\ref{lem:quasiconcavity}) is maximized over a 
convex, compact set.

\begin{lemma}[Concavity of $\Pi$]\label{lem:quasiconcavity}
Given $\pi^{+} \geq \pi^{-}$, the profit function $\Pi(\cdot)$ is concave in 
$\bm{w}^r$ and strictly concave in $\bm{w}^h$.
\end{lemma}
\begin{proof}
See Appendix~\ref{app:proofs}.
\end{proof}

\noindent Although the problem is concave, the objective function is 
non-differentiable due to the electricity payment function~\eqref{eq:ElectricPayment}.

\section{Optimal WDP Dispatch}\label{sec:Optimal}
We characterize here the optimal WDP dispatch and highlight its structural properties. Before presenting the optimal dispatch policy, we define the effective electricity prices of water production at the RODP's minimum and maximum limits as $\delta_{\mathsf{NZ}_1}:=f^{\prime r}(\underline{q}^r)\pi^w$ and
$\delta_{\mathsf{NZ}_2}:= f^{\prime r}(\overline{q}^r)\pi^w$, respectively. We first delineate the structure of the optimal dispatch under the standard condition where $ \pi^+ \geq \delta_{\mathsf{NZ}_1} \geq \delta_{\mathsf{NZ}_2} \geq \pi^-$. Cases violating the inequality are addressed in $\S$\ref{subsec:Special1}. Throughout this section we take the aggregate renewable generation $s := \mathbf{1}^\top \mathbf{g}$ as given, which corresponds to scheduling under a known renewable profile; the real-time setting, in which the plant is dispatched before $s$ is realized, is addressed in Appendix~\ref{app:MPC}.

\vspace{-0.2cm}
\subsection{Optimal WDP Dispatch Policy}\label{subsec:OptimalOperation}
To characterize the dispatch compactly, we define the {\em TDP stationarity function} $\widetilde{w}_t^h: \mathbb{R} \to [\underline{w}^h, \overline{w}^h]$ and the {\em aggregate controllable net consumption} $\Qc: \mathbb{R} \times \mathbb{R} \to \mathbb{R}$ as follows:
\begin{align}
    \widetilde{w}_t^h(\mu) &:= \left[ f_t^h\!\left(p_t^h(\mu)\right) \right]_{[\underline{w}^h,\, \overline{w}^h]}, \label{eq:wh_stationarity} \\
    \Qc(\mu, q^r) &:= T q^r - \sum_{t \in \mathcal{T}} \frac{\widetilde{w}_t^h(\mu)}{\eta_t^h}, \label{eq:AggPowerOp}
\end{align}
where $p_t^h(\mu) \in \mbbR_+$ is the unique solution to the implicit equation
\begin{equation}\label{eq:p_implicit}
    (C_t^h)'\!\left(p_t^h(\mu)\right) = f_t^{\prime h}\!\left(p_t^h(\mu)\right) \pi^w + \beta_t^h\!\left(p_t^h(\mu)\right) \mu.
\end{equation}
%\footnote{In the case of linear TDP fuel-to-water conversion function $f_t^h(p) = \alpha_t^h p$, equation~\eqref{eq:p_implicit} admits the closed-form solution $p_t^h(\mu) = D_t^h(\alpha_t^h \pi^w + (\alpha_t^h/\eta_t^h)\, \mu)$, recovering $\widetilde{w}_t^h(\mu) = \big[ \alpha_t^h\, D_t^h\!\big(\alpha_t^h \pi^w + (\alpha_t^h/\eta_t^h)\, \mu\big) \big]_{[\underline{w}^h,\, \overline{w}^h]}$.}
\begin{theorem}[Optimal WDP dispatch] \label{thm:optimal}
Assuming $ \pi^+ \geq \delta_{\mathsf{NZ}_1} \geq \delta_{\mathsf{NZ}_2} \geq \pi^-$, and given the aggregate renewable generation $s := \mathbf{1}^\top \mathbf{g}$, the optimal water dispatch $\{w_t^{h\ast}, w_t^{r\ast}\}$ obeys a four-threshold structure, with the four scalar thresholds $(\Gamma_{\mathsf{IM}}, \Gamma_{\mathsf{NZ}_1}, \Gamma_{\mathsf{NZ}_2}, \Gamma_{\mathsf{EX}})$ defined as:
\begin{equation} \label{eq:GammaDefs_new}
    \begin{aligned}
        \Gamma_{\mathsf{IM}} &:= \Qc(\pi^+, \underline{q}^r), \quad & \Gamma_{\mathsf{NZ}_1} &:= \Qc(\delta_{\mathsf{NZ}_1}, \underline{q}^r) \\
        \Gamma_{\mathsf{EX}} &:= \Qc(\pi^-, \overline{q}^r), \quad & \Gamma_{\mathsf{NZ}_2} &:= \Qc(\delta_{\mathsf{NZ}_2}, \overline{q}^r).
    \end{aligned}
\end{equation}
The optimal renewable-dependent dispatch is given in the table below:

\begin{table}[h]
\centering
\renewcommand{\arraystretch}{1.4}
\begin{tabular}{@{}llll@{}}
\toprule
\em{Mode} & \em{Range of} $s$ & \em{TDP} $w_t^{h\ast}(s)$ & \em{RODP} $w_t^{r\ast}(s)$ \\ \midrule
IM & $[0, \Gamma_{\mathsf{IM}})$ & $\widetilde{w}_t^h(\pi^+)$ & $\underline{w}^r$ \\
NZ & $[\Gamma_{\mathsf{IM}}, \Gamma_{\mathsf{NZ}_1})$ & $\widetilde{w}_t^h(\mu^{\mathsf{NZ}_1}(s))$ & $\underline{w}^r$ \\
NZ & $[\Gamma_{\mathsf{NZ}_1}, \Gamma_{\mathsf{NZ}_2}]$ & $\widetilde{w}_t^h(\mu^{\mathsf{NZ}}(s))$ & $f^r(\phi^r(\frac{\mu^{\mathsf{NZ}}(s)}{\pi^w}))$ \\
NZ & $(\Gamma_{\mathsf{NZ}_2}, \Gamma_{\mathsf{EX}}]$ & $\widetilde{w}_t^h(\mu^{\mathsf{NZ}_2}(s))$ & $\overline{w}^r$ \\
EX & $(\Gamma_{\mathsf{EX}}, \infty)$ & $\widetilde{w}_t^h(\pi^-)$ & $\overline{w}^r$ \\ \bottomrule
\end{tabular}
\end{table}
\noindent where $\mu^{\mathsf{NZ}_1}, \mu^{\mathsf{NZ}_2}$ are unique solutions to $\Qc(\mu, \underline{q}^r)=s$ and $\Qc(\mu, \overline{q}^r)=s$, respectively, and $\mu^{\mathsf{NZ}}$ solves $T\phi^r(\mu/\pi^w) - \sum_t \widetilde{w}_t^h(\mu)/\eta_t^h = s$.
\end{theorem}
\begin{proof}
See Appendix~\ref{app:proofs}.
\end{proof}
\noindent Although Theorem~\ref{thm:optimal} specifies the optimal 
water dispatch, it also implicitly schedules electricity consumption and 
production at each $t$, given the TDP and RODP conversion functions. The optimal setpoints of Theorem~\ref{thm:optimal} are depicted in Figure~\ref{fig:Optimal}.

\begin{figure}[t]
    \centering
    \resizebox{\columnwidth}{!}{%
        \input{Journal/tikZConcaveRO}%
    }
    \vspace{-0.65cm}
    \caption{Depiction of the optimal WDP dispatch in 
    Theorem~\ref{thm:optimal} and 
    Proposition~\ref{prop:SpecialCases}. Top row: Optimal TDP and RODP water dispatch (when 
    $\widetilde{w}_t^h(\pi^+)<\underline{w}^r$) versus aggregate renewables $s$. Bottom row: WDP's water and power 
    transactions with the water and electric utilities versus 
    $s$.}
    \label{fig:Optimal}
\end{figure}

\begin{corollary}[Optimal grid exchange]\label{prop:Optimalz}
The thresholds ($\Gamma_{\sf{IM}}, \Gamma_{\sf{EX}}$) in 
Theorem~\ref{thm:optimal} define the WDP's net-import mode with respect 
to the electric utility, \ie
\begin{equation}\label{eq:Optz}
\sgn(z^\ast(s)) = \begin{cases}
1 & ,\;s<\Gamma_{\sf{IM}} \\ 
0 & ,\; s \in [\Gamma_{\sf{IM}},\Gamma_{\sf{EX}}]\\
-1 & ,\;s>\Gamma_{\sf{EX}},
\end{cases}
\end{equation}
where $\sgn: \mbbR \rightarrow \{-1,0,1\}$.
\end{corollary}
\begin{proof}
See Appendix~\ref{app:proofs}.
\end{proof}

\noindent Corollary~\ref{prop:Optimalz} characterizes the WDP's value 
as a flexible grid resource (green curve in Figure~\ref{fig:Optimal}). The thresholds 
($\Gamma_{\sf{IM}}$, $\Gamma_{\sf{EX}}$) define three grid-interaction modes with the WDP providing:
\begin{itemize}[leftmargin=*]
    \item Controllable load (IM): aggregate net import 
    $z^\ast(s) > 0$ with demand response capability.
    \item Self-sufficient microgrid (NZ): net-zero operation $z^\ast(s) = 0$ matching generation with loads. 
    \item Dispatchable generator (EX): aggregate net export 
    $z^\ast(s) < 0$ with controllable output.
\end{itemize}
Critically, transitions between modes occur continuously through 
internal balancing of generation (TDP and renewables) and load (RODP), 
avoiding the discrete switching characteristic of conventional 
resources. The zero-crossing flexibility is particularly valuable for 
renewable integration, as the WDP can smoothly absorb forecast errors 
without mode changes. In the real-time setting, the policy of 
Theorem~\ref{thm:optimal} is applied in a receding-horizon manner that 
preserves the four-threshold structure (Appendix~\ref{app:MPC}).

%: the per-step problem reduces to 
%an instance of Theorem~\ref{thm:optimal} over the remaining horizon, so 
%its only approximation relative to the exact stochastic solution is the 
%renewable forecast error (Appendix~\ref{app:MPC}).
\vspace{-0.25cm}
\subsection{Optimal Dispatch Properties and Economical Insights}
The following properties characterize the structural properties of the optimal policy.

\begin{itemize}[leftmargin=*]
\item Offline threshold computation: The thresholds in \eqref{eq:GammaDefs_new} depend only on water/electricity tariff parameters and plant conversion functions. They are computed \textit{offline} in near-closed-form,\footnote{The thresholds are closed-form if $f_t^h(\cdot)$ is linear for all $t\in \Tc$.} independent of the realization of $\mathbf{g}$, offering a robust framework for real-time operation. A real-time receding-horizon policy that considers renewable uncertainty is presented in Appendix~\ref{app:MPC}.
\item Monotonicity and smoothing: The optimal TDP output $w_t^{h\ast}(s)$ is non-increasing in $s$, while $w_t^{r\ast}(s)$ is non-decreasing. Transitions between regions occur continuously via the dual variable of the net energy balance constraint.
\item RODP grid independence: Under the standard conditions of Theorem~\ref{thm:optimal}, the RODP dispatch is independent of the grid prices ($\pi^+,\pi^-$), being governed instead by the marginal revenue of RODP water relative to TDP costs.
\item Operational interdependence: In the NZ mode, the TDP and RODP interchangeably use the available renewables to maintain energy-balancedness over the billing period.
\end{itemize}

The off-grid mode $s \in [\Gamma_{\mathsf{IM}}, \Gamma_{\mathsf{EX}}]$ (see green curve in Figure~ \ref{fig:Optimal}) illustrates the WDP's internal flexibility. In this regime, revenue arises solely from water sales, and the WDP maintains a net-zero electricity balance. When $s\in [\Gamma_{\mathsf{IM}},\Gamma_{\mathsf{NZ}_1})$, as renewables increase, the RODP remains at its minimum while renewables displace expensive TDP generation. At $s = \Gamma_{\mathsf{NZ}_1}$, the TDP marginal cost equals the marginal revenue of RODP water ($\delta_{\mathsf{NZ}_1}$). When $s\in [\Gamma_{\mathsf{NZ}_1},\Gamma_{\mathsf{NZ}_2}]$, both assets adjust simultaneously: additional renewables are used to both increase RODP production and further reduce TDP output, optimized via $\mu^{\mathsf{NZ}}$. When $s\in [\Gamma_{\mathsf{NZ}_2},\Gamma_{\mathsf{EX}}]$, the RODP is saturated at $\overline{w}^r$, and further renewable increases continue to displace TDP output until $s$ reaches $\Gamma_{\mathsf{EX}}$, at which point exporting becomes the more profitable strategy.

\subsection{Special Tariff Cases}\label{subsec:Special1}
When the standard condition of Theorem~\ref{thm:optimal} fails because $\pi^-$ and $\pi^+$ both lie outside the interval $[\delta_{\mathsf{NZ}_2}, \delta_{\mathsf{NZ}_1}]$, the RODP saturates at a bound throughout, as follows.

\begin{proposition}[Optimal dispatch under imbalanced tariffs]
\label{prop:SpecialCases}
When both grid prices lie outside of $[\delta_{\mathsf{NZ}_2}, \delta_{\mathsf{NZ}_1}]$, \ie $\pi^-,\pi^+ \notin [\delta_{\mathsf{NZ}_2}, \delta_{\mathsf{NZ}_1}]$, the RODP dispatch is static:
\begin{equation}\label{eq:OptWrSpecial}
w_t^{r\ast} = \begin{cases}
\overline{w}^r & ,\; \pi^-\leq\pi^+<\delta_{\mathsf{NZ}_2}\leq\delta_{\mathsf{NZ}_1} \\
\underline{w}^r & ,\; \delta_{\mathsf{NZ}_2}\leq\delta_{\mathsf{NZ}_1}<\pi^-\leq\pi^+,
\end{cases}
\end{equation}
and the TDP follows a two-threshold policy with $\Gamma_{\sf{IM}}^{\mathsf{s}} := \Qc(\pi^+, (f^r)^{-1}(w^{r\ast}))$ and $\Gamma_{\sf{EX}}^{\mathsf{s}} := \Qc(\pi^-, (f^r)^{-1}(w^{r\ast}))$:
\begin{equation}\label{eq:OptWhSpecial}
w_t^{h\ast}(s) = \begin{cases}
\widetilde{w}_t^h(\pi^+) & ,\; s < \Gamma_{\sf{IM}}^{\mathsf{s}} \\
\widetilde{w}_t^h(\mu(s)) & ,\; s \in [\Gamma_{\sf{IM}}^{\mathsf{s}},\Gamma_{\sf{EX}}^{\mathsf{s}}] \\
\widetilde{w}_t^h(\pi^-) & ,\; s > \Gamma_{\sf{EX}}^{\mathsf{s}},
\end{cases}
\end{equation}
where $\mu(s)\in [\pi^-,\pi^+]$ solves $\Qc(\mu, (f^r)^{-1}(w^{r\ast})) = s$.
\end{proposition}
\begin{proof}
See Appendix~\ref{app:proofs}.
\end{proof}

\noindent In words, Proposition~\ref{prop:SpecialCases} says that when both grid prices exceed the RODP's marginal water value over its entire range, buying electricity to desalinate is not worth it and the RODP idles at $\underline{w}^r$; when both fall below it, electricity is always cheap enough that the RODP saturates at $\overline{w}^r$. Either way the RODP's operation is independent of $s$, leaving the TDP to follow $s$ based on a two threshold rule. When a grid price lies inside $(\delta_{\mathsf{NZ}_2}, \delta_{\mathsf{NZ}_1})$, the RODP is interior at that price, and the two-threshold rule above generalizes to a three- or two-threshold rule, depending on which of $\pi^+,\pi^-$ lies inside $(\delta_{\mathsf{NZ}_2},\delta_{\mathsf{NZ}_1})$. We state this extended case in Appendix~\ref{app:proofs}.

\section{Special Cases and Profit Sensitivity}\label{sec:SpecialCase}
% =====================================================================
%  Section IV — SPECIAL TECHNOLOGY CASES AND PROFIT SENSITIVITY
%====================================================================
We discuss in $\S$\ref{subsec:Special2} special cases arising from the WDP's available desalination technology and in $\S$\ref{subsec:profits} the WDP's profitability and sensitivity under optimal dispatch.

\subsection{Special Technology Cases: RODP-Only and TDP-Only}
\label{subsec:Special2}
The optimal dispatch for WDPs equipped with only a single desalination technology is formalized in Corollary \ref{corol:RODPTDPonly}.

\begin{corollary}[Optimal dispatch under single desalination technology]
\label{corol:RODPTDPonly}
Assuming $ \pi^+ \geq \delta_{\mathsf{NZ}_1} \geq \delta_{\mathsf{NZ}_2} \geq \pi^-$ and given the aggregate renewable generation $s := \mathbf{1}^\top \mathbf{g}$, the optimal dispatch of an RODP-only WDP  at each $t \in \mathcal{T}$ is a two-threshold policy, given by
\begin{equation}\label{eq:WrOnly}
w_t^{r\ast}(s) = \begin{cases}
\underline{w}^r & ,\; s < T\,\underline{q}^r \\ 
f^r(\phi^r(\frac{\mu^{\mathsf{NZ}}(s)}{\pi^w})) & ,\; s \in [T\,\underline{q}^r,\, T\,\overline{q}^r] \\
\overline{w}^r & ,\; s > T\,\overline{q}^r,
\end{cases}
\end{equation}
where $\mu^{\mathsf{NZ}}(s)$ is the solution of $T\phi^r(\mu/\pi^w) = s$.

For a TDP-only WDP, the optimal dispatch at each $t \in \mathcal{T}$ is independent of $s$ and is given by $w_t^{h\ast} = \widetilde{w}_t^h(\pi^-)$.
\end{corollary}

\begin{proof}
See Appendix~\ref{app:proofs}.
\end{proof}

In the RODP-only case, the WDP acts as a controllable load under net-metering, leveraging the billing period to arbitrage energy consumption against water demand. In the TDP-only case, the WDP is a power producer, and since there is no local load to offset, the optimal setpoint is determined solely by the export price $\pi^-$ and water price $\pi^w$.

\vspace{-0.25cm}
\subsection{WDP Profits under Optimal Dispatch}\label{subsec:profits}
The optimal WDP profit $\Pi^\ast(s)$ increases with the aggregate renewables $s$, but at a non-increasing rate. This concavity, together with the profit's sensitivity to the tariff and technology parameters, is made precise next.

\begin{theorem}[Operation profit sensitivity analysis]\label{thm:maxprofit}
Assuming $ \pi^+ \geq \delta_{\mathsf{NZ}_1} \geq \delta_{\mathsf{NZ}_2} \geq \pi^-$, and given the aggregate renewables $s := \mathbf{1}^\top \mathbf{g}$, the maximum profit $\Pi^\ast(s)$ increases with $s$ at a marginal rate $d\Pi^\ast/ds$ characterized by the following regimes:
\begin{itemize}[leftmargin=*]
    \item \textbf{IM Regime ($s < \Gamma_{\mathsf{IM}}$):} The profit increases at the rate of the import price, $\pi^+$.
    \item \textbf{NZ Regime (boundary) ($s \in [\Gamma_{\mathsf{IM}}, \Gamma_{\mathsf{NZ}_1})$):} The rate $x$ decreases continuously such that $x \in [\delta_{\mathsf{NZ}_1}, \pi^+]$.
    \item \textbf{NZ Regime (interior) ($s \in [\Gamma_{\mathsf{NZ}_1}, \Gamma_{\mathsf{NZ}_2}]$):} The rate $x$ decreases continuously such that $x \in [\delta_{\mathsf{NZ}_2}, \delta_{\mathsf{NZ}_1}]$.
    \item \textbf{NZ Regime (boundary) ($s \in (\Gamma_{\mathsf{NZ}_2}, \Gamma_{\mathsf{EX}}]$):} The rate $x$ decreases continuously such that $x \in [\pi^-, \delta_{\mathsf{NZ}_2}]$.
    \item \textbf{EX Regime ($s > \Gamma_{\mathsf{EX}}$):} The profit increases at the rate of the export price, $\pi^-$.
\end{itemize}
Moreover, $\Pi^\ast(s)$ is: (i) non-increasing in $\pi^+$, (ii) non-decreasing in $\pi^-$, (iii) non-decreasing in $\pi^w$, (iv) non-decreasing in $f^r$ (pointwise), and (v) non-decreasing in $f_t^h$ (pointwise) $\forall t \in \mathcal{T}$.
\end{theorem}

\begin{proof}
See Appendix~\ref{app:proofs}.
\end{proof}

The diminishing marginal profit with respect to $s$ reflects the shifting value of renewables across operational modes. In the IM mode, renewables directly offset grid purchases at the import price $\pi^+$. In the EX mode, they are sold at the export price $\pi^-$. Within the NZ region, the marginal value of a renewable unit is endogenously determined by the {\em avoided cost} of TDP production or the {\em marginal revenue} of increased RODP production, both of which are bounded by $[\pi^-, \pi^+]$.

%As shown in Figure~\ref{fig:Optimal}, while electricity payments and TDP costs decrease monotonically with $s$, the water revenue is non-monotonic, as it depends on the internal balance between the TDP and RODP within the NZ regime.

%\begin{comment}
    \section{Numerical Results}\label{sec:num}
To showcase the optimal dispatch of the WDP and the effect of tariff and plant parameters, we consider a desalination plant with natural-gas-powered thermal units, RO units, and solar PV renewables operating within a competitive desalinated-water market. The plant sells water to a water utility, and transacts with an electric utility based on its net electricity exchange. The water and electricity tariff periods are assumed to be hourly, while the plant dispatch period is 10 minutes. 

While we do sensitivity analysis on the effect of the plant parameters, we consider a base case with realistic parameter settings. In the base case, the water price is $\pi^w =
\$1/m^3$,\footnote{This is within the price range of potable water in Saudi Arabia. See \href{https://www.marafiq.com.sa/en/partnering-with-us/water-tariff/}{Marafiq}.} and the electricity is priced based on a NEM tariff
with a ToU retail rate of $\pi^+_{\mbox{\tiny ON}} = \$400$/MWh and $\pi^+_{\mbox{\tiny OFF}} = \$200$/MWh peak and offpeak prices and 16--21 peak period, and a sell rate of $\pi^- = \$100$/MWh.

\begin{figure}[t]
    \centering
    \includegraphics[width=0.85\linewidth]{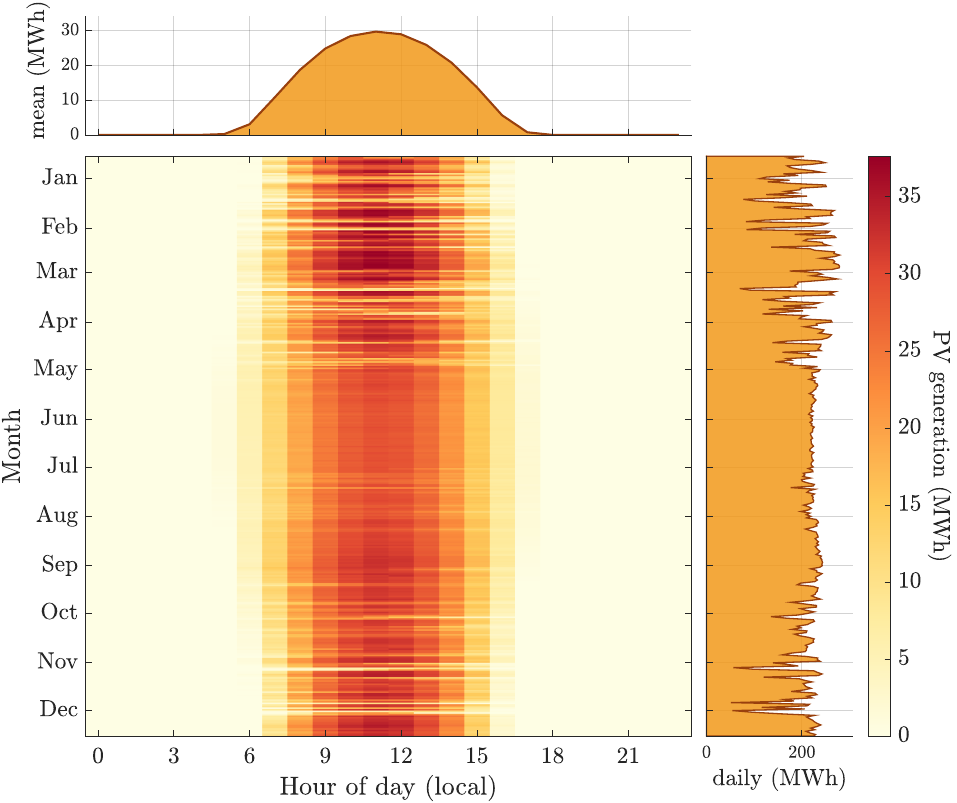}
    \vspace{-0.3cm}
    \caption{Hourly PV generation of the 45~MW Jubail array over 2019 (main panel), with the mean daily profile (top) and the daily energy (right).}
    \label{fig:WDPNum0}
\end{figure}

The plant's PV capacity was set to 45~MW so that peak solar output (37.6~MWh) matches the plant's peak reverse-osmosis demand (35.7~MWh), placing the plant near the import-export boundary at midday so the dispatch extends over all operating modes. One-year (2019) hourly PV generation for Jubail city in Saudi Arabia was obtained from \cite{Pfenninger2016}, which offers a satellite-derived solar irradiance data. Figure~\ref{fig:WDPNum0} shows the hourly generation over the year together with its mean daily profile and daily energy. The PV plant produced 77{,}276~MWh in 2019 with a peak hourly output of 37.6~MWh. Notably, the midday output is nearly season-invariant; the monthly mean noon generation varies only between 26.6 and 31.0~MWh.

We used a convex quadratic and time-invariant TDP operating cost function $C_h(p_h) = a p^2_h + b p_h + c$, with $a = \$0.008/\text{MBTU}^2$, $b = \$4$/MBTU, and $c =\$0$ \cite{Santhosh&Farid&Toumi:14AE}. The concave-quadratic conversion functions were parameterized as $f_h(p_h)=\alpha_h p_h - \frac{1}{2}\rho_h p^2_h$, with $\alpha_h = 5\; m^3$/MBTU and $\rho_h = 0.00213\; m^3/\text{MBTU}^2$, and $f_r(q_r)=\alpha_r q_r - \frac{1}{2}\rho_r q^2_r$, with $\alpha_r = 300\; m^3$/MWh and $\rho_r = 3.73\; m^3/\text{MWh}^2$ \cite{Elimelech:2011Science}. The water-to-electricity co-production ratio of the TDP was fixed at $\eta_h = 90\; m^3$/MWh, reflecting an RO-dominant hybrid plant. All of the parameters are varied in $\S$\ref{subsec:NumSensitivity}. The RODP and TDP flow limits were set to $\overline{w}_r = 8{,}333\; m^3/h , \overline{w}_h = 3{,}000\; m^3/h$.

\vspace{-0.1cm}
\subsection{WDP Dispatch Algorithms}
We compare the optimal WDP dispatch in Theorem \ref{thm:optimal} and Proposition \ref{prop:SpecialCases} to two other standard dispatch algorithms. To ensure a fair comparison, all modes have the same resources; additionally, they all face the same water and electricity tariffs.

\subsubsection{Max-RODP Algorithm}
Under this case, the WDP sets the RODP to its maximum
$w_t^r = \overline{w}^r, \forall t \in \Tc$ and co-optimizes the
renewables with the TDP, therefore, the optimal TDP dispatch becomes a
special case of Theorem~\ref{thm:optimal} and
Proposition~\ref{prop:SpecialCases}, given by
\begin{equation*}
w_t^{h\ast}(\bm{g}) = \begin{cases}
\widetilde{w}_{t}^h(\pi^+) & ,\;s<\Qc(\pi^+,\overline{q}^r) \\
\widetilde{w}_t^h(\mu^{\mathsf{NZ}}(s))
& ,\; s \in [\Qc(\pi^+,\overline{q}^r),\Qc(\pi^-,\overline{q}^r)]\\
\widetilde{w}_{t}^h(\pi^-) & ,\;s>\Qc(\pi^-,\overline{q}^r),
\end{cases}
\end{equation*}
where
$\Qc(\cdot)$ and $\widetilde{w}_{t}^h(\mu)$ are as in~\eqref{eq:wh_stationarity}-\eqref{eq:AggPowerOp} and the price $\mu^{\mathsf{NZ}}(s) \in [\pi^-,\pi^+]$ solves $T\,\overline{q}^r - \sum_t \widetilde{w}_t^h(\mu)/\eta_t^h = s$.

\subsubsection{Passive-TDP Algorithm}
Here, the TDP is dispatched as if the plant were an independent water
and power producer (IWPP); therefore, it always faces $\pi^-$. As per Corollary~\ref{corol:RODPTDPonly}, the optimal TDP dispatch becomes $w_t^{h\ast} = \widetilde{w}_t^h(\pi^-)$ for all $t \in \Tc$, and as a result, from Theorem~\ref{thm:optimal}, $\Gamma_{\sf{IM}} = \Gamma_{\sf{NZ}_1}$ and $\Gamma_{\sf{EX}} = \Gamma_{\sf{NZ}_2}$. The optimal RODP schedule becomes a two-threshold policy.

\vspace{-0.25cm}
\subsection{WDP Daily Dispatch}
\begin{figure*}[t]
    \centering
    \includegraphics[width=0.98\linewidth]{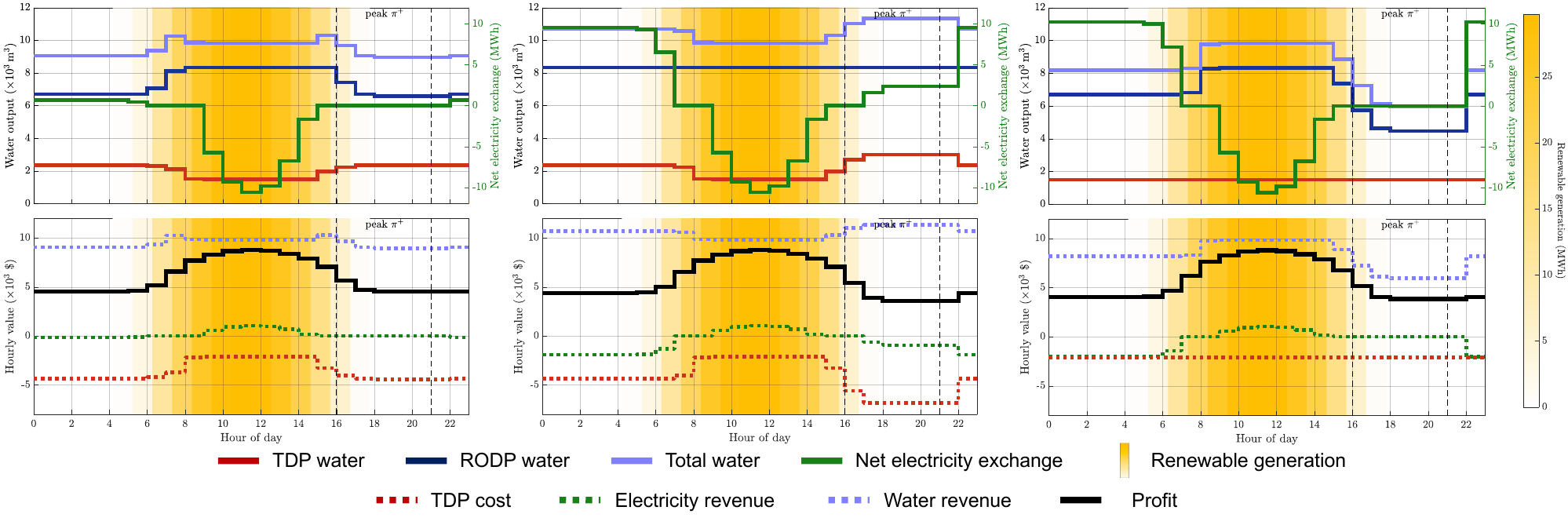}
    \vspace{-0.45cm}
    \caption{WDP hourly dispatch and profits under the optimal joint dispatching (left), max-RODP algorithm (center), and passive-TDP algorithm (right).}
    \label{fig:WDPNum1}
\end{figure*}

Figure \ref{fig:WDPNum1} shows the WDP's optimal water output and net electricity exchange (top) and profits (bottom) under the three dispatch algorithms considered over an average PV day profile.

Under optimal dispatch, the WDP remained nearly self-sufficient throughout the day. During offpeak hours without solar generation, the offpeak import price lies strictly inside the net-zero threshold band, $\pi^- < \pi^+_{\mbox{\tiny OFF}} < f_r'(0)\,\pi^w = \$300$/MWh, so the RODP was dispatched at the interior setpoint at which its marginal water value equals the import price, yielding about 2{,}350~m\(^3\) (TDP) and 6{,}702~m\(^3\) (RODP); the TDP co-generation covered all but a residual 0.7~MWh of the RODP load. As renewables increased, the plant entered the net-zero regime and allocated the additional energy to the RODP, whose marginal water value exceeded the value of exporting electricity, raising RODP output to 7{,}059~m\(^3\) in hour~6 and 8{,}130~m\(^3\) in hour~7. By hour~8, renewables drove the RODP to its maximum of 8{,}333~m\(^3\), reducing reliance on the TDP.

Between hours~9 and~14, high renewable output pushed the WDP into net-export mode, forcing the TDP to its export setpoint $\tilde{w}^h(\pi^-)$ (1{,}490~m\(^3\)) because of the low export price~\(\pi^-\), while the RODP remained at maximum. Although electricity-export revenue was small, hourly profits nearly doubled (from \$4{,}548 to \$8{,}784) due to higher water sales and lower TDP costs. During the evening peak window, importing at $\pi^+_{\mbox{\tiny ON}} > f_r'(0)\,\pi^w$ was unattractive, hence the TDP generation increased to fully feed RODP and avoid peak-priced imports.

Under the \textit{max-RODP} algorithm, only the TDP adjusted to renewable availability, causing the WDP to import 9.6~MWh in every solar-free offpeak hour and, with the TDP driven to its maximum of 3{,}000~m\(^3\) during the evening peak at the high rate $\pi^+_{\mbox{\tiny ON}}$. Profitability therefore declined, with profits near \$4{,}400 during these periods despite higher water-sales revenue.

Under the \textit{passive-TDP} algorithm, the TDP produced a fixed 1{,}490~m\(^3\)/h, incurring a constant operating cost of \$2{,}097. The plant imported 10.3~MWh in solar-free hours, while during the evening peak the RODP was throttled to about 4{,}456~m\(^3\), matching the TDP co-generation, to avoid peak-priced imports. Profits closely followed the water-revenue profile, as TDP costs were fixed and electricity-export revenue remained small because most renewables were consumed by the RODP. Total daily profits under \textit{joint dispatching}, \textit{max-RODP}, and \textit{passive-TDP} were \$142{,}158, \$135{,}886, and \$132{,}996, respectively.

\begin{figure}[t]
    \centering
    \includegraphics[width=\linewidth]{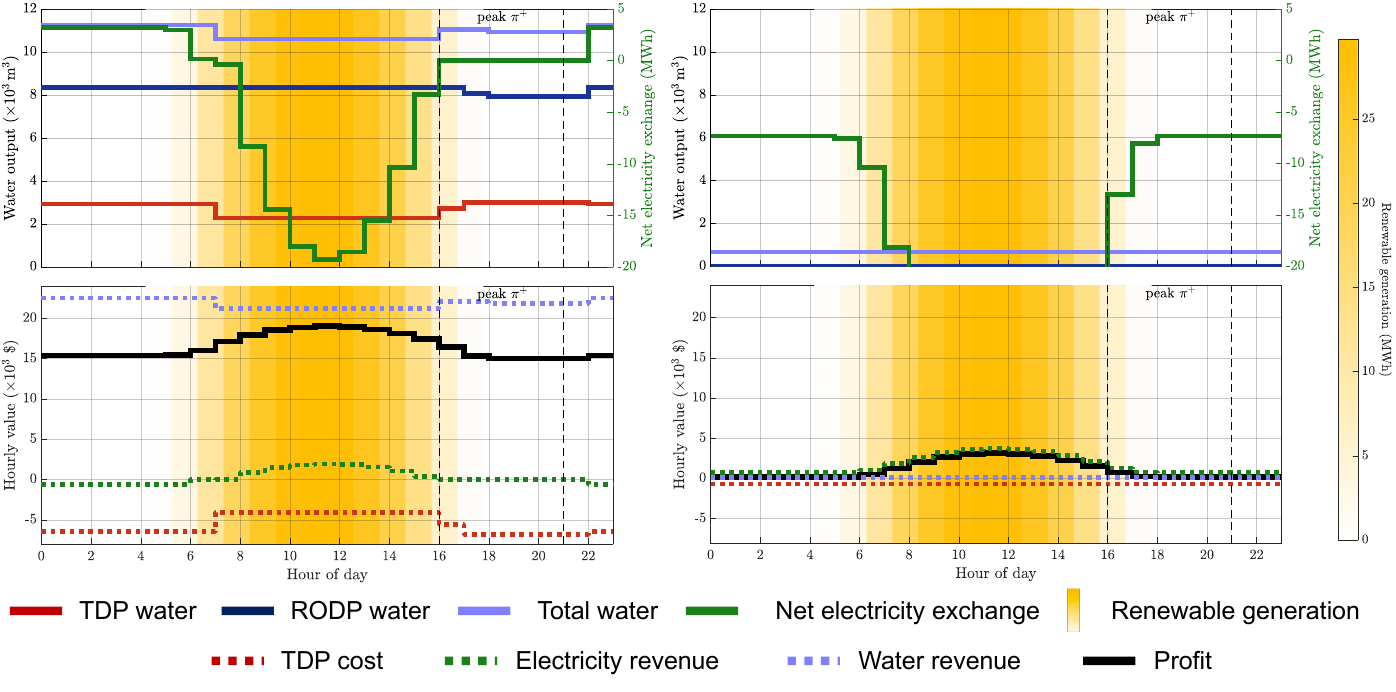}
    \vspace{-0.4cm}
    \caption{WDP hourly dispatch and profits under the optimal policy, when $\pi^w=\$2/m^3$ (left) and $\pi^w=\$0.2/m^3$ (right).}
    \label{fig:WDPNum2}
\end{figure}

Figure~\ref{fig:WDPNum2} presents the hourly optimal WDP dispatch and profits for two water-price scenarios: a high price of $\pi^w=\$2/m^3$ (left) and a low price of $\pi^w=\$0.2/m^3$ (right). Under high water price, the RODP operated at its maximum of 8{,}333~m$^3$ outside the peak period. During offpeak hours without renewable generation, this output was supported by grid imports of around 3.2~MWh, since the marginal revenue from water sales exceeds the marginal cost of electricity imports. During peak, the plant settled at net zero with the TDP at its maximum. The resulting total daily profit was \$395{,}668.

When $\pi^w$ was low, the RODP was off because exporting electricity is more profitable than producing water, effectively turning the plant into an IWPP. The TDP water remained constant at 660m$^3$/h, as the TDP operated at $\tilde{w}^h(\pi^-)$. Electricity sales (\$38{,}778) were much higher than water revenue (\$3{,}169), and the total daily profit dropped to \$25{,}342.

\vspace{-0.3cm}
\subsection{Net Electricity Exchange}
\begin{figure}[t]
    \centering
    \includegraphics[width=\linewidth]{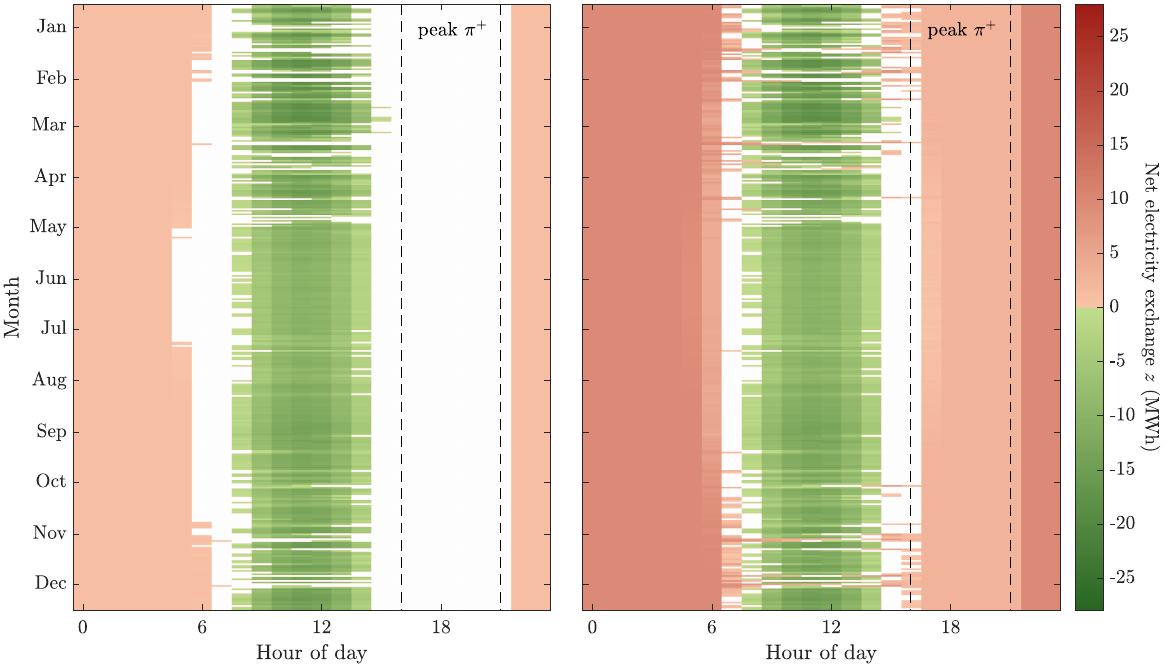}
    \vspace{-0.3cm}
    \caption{Net electricity exchange $z$ over the year under the optimal dispatch (left) and the max-RODP algorithm (right). White cells denote net-zero.}
    \label{fig:WDPNum3}
\end{figure}

Figure~\ref{fig:WDPNum3} maps the net electricity exchange $z$ over the year under the optimal and max-RODP algorithms. Two observations stand out. First, under the optimal dispatch, the WDP was mostly net-balanced or exporting year-round: 41\% of all hours were at $z=0$, and total imports over the year was only 2{,}027 MWh. The max-RODP plant, in contrast, imported 35{,}049 MWh. Both algorithms exported identically at roughly hours 9-15, (17{,}594 MWh annually), since both operated the RODP at maximum with the TDP at its export setpoint $\tilde{w}^h(\pi^-)$. Second, the within-day structure is highly season-invariant: the import/net-zero/export pattern of Thm.~\ref{thm:optimal} holds in every month, a direct consequence of the flat renewable seasonal profile in Fig.~\ref{fig:WDPNum0}.

\vspace{-0.3cm}
\subsection{Sensitivity of Profit to Tariff and Technology Parameters}\label{subsec:NumSensitivity}
\begin{figure}[t]
    \centering
    \includegraphics[width=0.95\linewidth]{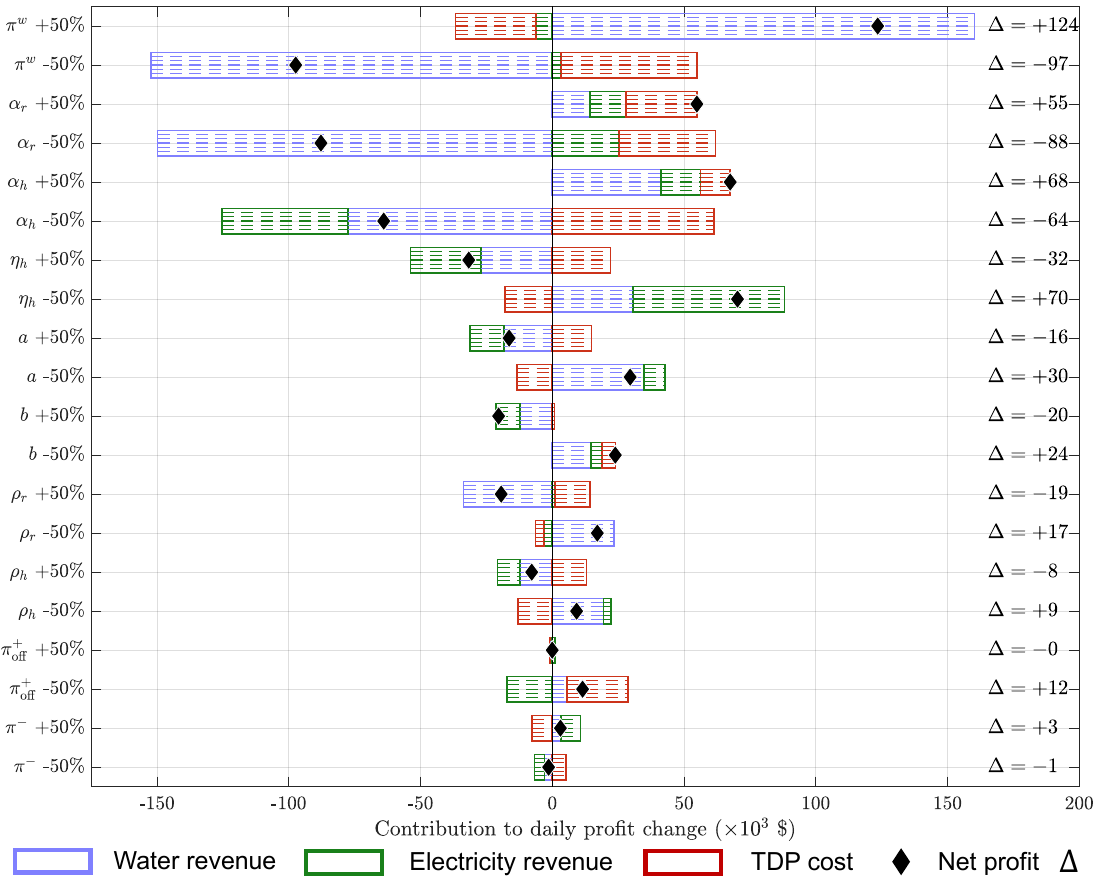}
    \vspace{-0.3cm}
    \caption{Sensitivity of WDP daily profit to $\pm$50\% variations in tariff and technology parameters, decomposed into water revenue, electricity revenue, and TDP cost. Rows are ordered by the profit swing.}
    \label{fig:WDPNum4}
\end{figure}

Figure~\ref{fig:WDPNum4} decomposes the average daily profit change under optimal dispatch to TDP fuel cost, and electricity and water revenues due to $\pm 50\%$ change in the tariff and technology parameters. 

A $50\%$ increase in the water price $\pi^{w}$ produced the largest positive effect, raising daily profit by approximately \$123.5k despite the higher TDP cost, because the increase in water revenue (+\$160.2k) dominated all other effects. Conversely, reducing $\pi^{w}$ by $50\%$ resulted in a profit loss of about \$97.4k.

Increasing the RO factor $\alpha_{r}$ by $50\%$ raised profit by \$54.9k, with the gain split between higher water sales and, because a more efficient RODP frees TDP generated electricity, lowering both TDP and grid imports costs. Symmetrically, reducing $\alpha_{r}$ by $50\%$ cost \$87.8k, dominated by the cut in RO water revenue that was only partially offset by the TDP cost reduction as it operated at the export setpoint $\tilde{w}^h(\pi^-)$. The co-production ratio $\eta_{h}$ acted in the opposite direction: reducing it by $50\%$ (a more power-rich TDP) raised profit by \$70.3k, mainly through electricity revenue (+\$57.5k). The parameters $\rho_{r}$ and $\rho_{h}$ contributed changes of \$36.5k and \$17.0k, respectively.

Changes in the offpeak import price $\pi^{+}_{\mbox{\tiny OFF}}$ had a negligible effect when increased by $50\%$ as the higher price pushed imports to zero. A $50\%$ decrease produced a small profit increase (about \$11.5k), however, it reveals an interesting coupling between electricity and water dispatch. Lower import prices incentivize increased RO dispatch, which increases RO water output and electricity consumption, raising water-sales revenue while allowing the TDP to operate at its export setpoint $\tilde{w}^h(\pi^-)$ and cut fuel cost by \$23.1k at the expense of \$17.2k in electricity revenue. A similar coupling appears for the export price $\pi^{-}$: as $\pi^{-}$ increased, exporting electricity became more profitable, increasing the incentive to run TDP, which resulted in higher water and electricity revenues at a higher fuel cost.

\vspace{-0.6cm}
\subsection{Optimality Gap of the Max-RODP Algorithm}
\begin{figure}[t]
    \centering
    \includegraphics[width=\linewidth]{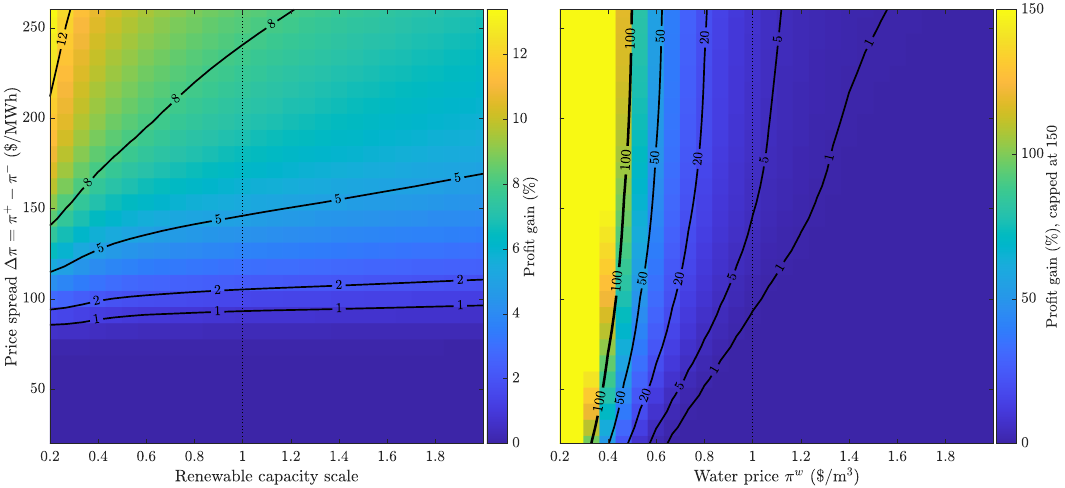}
    \vspace{-0.7cm}
    \caption{Profit gain (\%) of the optimal policy over max-RODP algorithm, versus the price spread $\Delta\pi=\pi^{+}-\pi^{-}$ and (left) renewable capacity, or (right) water price $\pi^{w}$. Dotted lines mark baseline values.}
    \label{fig:WDPNum5}
\end{figure}

Figure~\ref{fig:WDPNum5} reports the profit gain of the optimal dispatch over the standard max-RODP algorithm, against the NEM price differential $\Delta\pi=\pi^{+}-\pi^{-}$ (with $\pi^{-}=\$100$/MWh) and, respectively, (left)~the renewable capacity scale, or (right)~the water price $\pi^{w}$; the dotted line marks the baseline value in each panel. Renewable capacity alone had a modest effect: the gap stayed below about $13\%$ and widened only gradually with the price spread. The water price is far more decisive: the gap collapsed to zero once $\pi^{w}$ reaches roughly $\$1/m^{3}$, where maximizing RO output is itself optimal, but increased sharply as $\pi^{w}$ dropped, crossing $100\%$ (bold contour) at a higher water price as the spread widened, the point past which max-RODP operates at a loss by producing RO water whose marginal value is below the cost of the energy it consumes.

%\end{comment}

\section{Conclusion}\label{sec:conclusion}
 We present a tractable analytical solution to the joint water-power dispatching problem for WDPs operating as hybrid generator-load resources, yielding valuable operational and economical insights. The optimal dispatch exhibits a threshold-based structure determined solely by technology and tariff parameters, and in some modes is independent of electricity prices. The coupled optimal dispatch of thermal and RO units is a function of local renewables, and is divided into five regions. When renewable generation is very low or very high, the WDP maintains constant water production while using renewables to offset electricity imports or reduce exports. In the three intermediate modes, the WDP is energy-balanced, and the surplus renewables are allocated to water production: first by reducing thermal output, then by increasing RO output up to capacity, and finally by further displacing thermal generation when its marginal cost exceeds the electricity export price.

The structural insights and performance benefits demonstrated through simulations showcase the potential of the efficient dispatch of integrated water-energy systems in increasingly renewable-rich and market-responsive environments.

To handle renewable uncertainty, we develop an MPC-based dispatch algorithm that preserves the threshold structure, yielding a tractable real-time implementation that avoids the complexity of the underlying multistage stochastic problem.

Several avenues for future research follow from the limitations of this work. First, incorporating energy storage would enable temporal shifting of desalinated water and electricity. Second, our analysis focuses on short-run operations and abstracts from long-term investment decisions. Extending the framework to a joint planning-operation model would inform optimal sizing of desalination technologies and renewable capacity. Third, WDPs may provide additional value streams to electrical grids, including frequency regulation or generation capacity service, which are not captured in our model.

% Deleted:
%Finally, an interesting future direction involves modeling the interaction between the thermal and RO-based units as a cooperative game. Such a game-theoretic formulation could reveal strategic behaviors, shared incentives, and fair resource allocations between technologies under different market and operational scenarios.

%\section*{Acknowledgment}
%The authors are grateful for the insights and discussions with Prof. X from X University.
\vspace{-0.14cm}
{
\bibliographystyle{IEEEtranDOI}
\bibliography{WattsDrops}

@ARTICLE{Alahmed&Tong:22IEEETSG,  author={Alahmed, Ahmed S. and Tong, Lang},  journal={IEEE Transactions on Smart Grid},   title={On Net Energy Metering {X}: Optimal Prosumer Decisions, Social Welfare, and Cross-Subsidies},   year={2023},  volume={14},  number={02},  doi={10.1109/TSG.2022.3158951}}

@techreport{TurningTheTide:23GlobalWaterCommission,
	title = {Turning the Tide: A Call to Collective Action},
	year = {2023},
	month = {03},
	journal = {Global Commission on the Economics of Water},
author ={{Global Commission on the Economics of Water}},
    url = {https://watercommission.org/wp-content/uploads/2023/03/Turning-the-Tide-Report-Web.pdf},
institution ={Global Commission on the Economics of Water}
}

@book{Boyd&Vandenberghe:04Book,
  title={Convex Optimization},
  author={Boyd, Stephen and Vandenberghe, Lieven},
  year={2004},
  doi ={
https://doi.org/10.1017/CBO9780511804441},
  publisher={Cambridge University Press}
}

@article{Elimelech:2011Science,
  author    = {Menachem Elimelech and William A. Phillip},
  title     = {The Future of Seawater Desalination: Energy, Technology, and the Environment},
  journal   = {Science},
  volume    = {333},
  number    = {6043},
  pages     = {712--717},
  year      = {2011},
  doi       = {10.1126/science.1200488}
}

@article{Ghaffour:2015Desalination,
  author    = {Noreddine Ghaffour and Jochen Bundschuh and Hassan Mahmoudi and Mohammed F. A. Goosen},
  title     = {Renewable energy-driven desalination technologies: A comprehensive review on challenges and potential applications},
  journal   = {Desalination},
  volume    = {356},
  pages     = {94--114},
  year      = {2015},
  doi       = {10.1016/j.desal.2014.10.024}
}

@ARTICLE{ONeiletal:24OAJPE,
  author={O’Neil, Rebecca and Oikonomou, Konstantinos and Tidwell, Vince and Voisin, Nathalie and Kerby, Jessica and Jason Hou, Z. and Parvania, Masood and Al-Awami, Ali T. and Panteli, Mathaios and Conrad, Steven A. and Brekken, Ted K. A.},
  journal={IEEE Open Access Journal of Power and Energy}, 
  title={Global Research Priorities for Holistic Integration of Water and Power Systems}, 
  year={2024},
  volume={11},
  number={},
  pages={457-468},
  doi={10.1109/OAJPE.2024.3457448}}

@article{Santhosh&Farid&Toumi:14AE,
  author    = {Apoorva Santhosh and Amro M. Farid and Kamal Youcef-Toumi},
  title     = {Real-time economic dispatch for the supply side of the energy-water nexus},
  journal   = {Applied Energy},
  volume    = {122},
  pages     = {42--52},
  year      = {2014},
  issn = {0306-2619},
  doi = {https://doi.org/10.1016/j.apenergy.2014.01.062},
}

@article{Ghaithan2022,
  author    = {A. M. Ghaithan and A. Mohammed and A. Al-Hanbali and A. M. Attia and H. Saleh},
  title     = {Multi-objective optimization of a photovoltaic-wind-grid connected system to power reverse osmosis desalination plant},
  journal   = {Energy},
  volume    = {251},
  pages     = {123888},
  year      = {2022},
  issn = {0360-5442},
doi = {https://doi.org/10.1016/j.energy.2022.123888}
}

@article{Qudah2024,
  author    = {A. Qudah and A. Almerbati and E. M. Mokheimer},
  title     = {Novel approach for optimizing wind-pv hybrid system for {RO} desalination using differential evolution algorithm},
  journal   = {Energy Conversion and Management},
  volume    = {300},
  pages     = {117949},
  issn = {0196-8904},
doi = {https://doi.org/10.1016/j.enconman.2023.117949},
  year      = {2024}
}

@article{MoazeniAPEN2020,
  author    = {F. Moazeni and J. Khazaei and J. P. Pera Mendes},
  title     = {Maximizing energy efficiency of islanded micro water-energy nexus using co-optimization of water demand and energy consumption},
  journal   = {Applied Energy},
  volume    = {266},
  pages     = {114863},
  issn = {0306-2619},
doi = {https://doi.org/10.1016/j.apenergy.2020.114863},
  year      = {2020}
}

@ARTICLE{AlAwamiTPS2022,
  author={Elsir, Mohamed and Al-Awami, Ali T. and Antar, Mohamed A. and Oikonomou, Konstantinos and Parvania, Masood},
  journal={IEEE Transactions on Power Systems}, 
  title={Risk-Based Operation Coordination of Water Desalination and Renewable-Rich Power Systems}, 
  year={2023},
  volume={38},
  number={2},
  pages={1162-1175},
  doi={10.1109/TPWRS.2022.3174565}}

@ARTICLE{Oikonomou&Parvania:20TSG,
  author={Oikonomou, Konstantinos and Parvania, Masood},
  journal={IEEE Transactions on Smart Grid}, 
  title={Optimal Coordinated Operation of Interdependent Power and Water Distribution Systems}, 
  year={2020},
  volume={11},
  number={6},
  pages={4784-4794},
  doi={10.1109/TSG.2020.3000173}}

@ARTICLE{Almehizia&Almasri&Hussein&Ehsani:19TSG,
  author={Almehizia, Abdullah A. and Al-Masri, Hussein M. K. and Ehsani, Mehrdad},
  journal={IEEE Transactions on Smart Grid}, 
  title={Integration of Renewable Energy Sources by Load Shifting and Utilizing Value Storage}, 
  year={2019},
  volume={10},
  number={5},
  pages={4974-4984},
  doi={10.1109/TSG.2018.2871806}}

@ARTICLE{Guoetal:16TSG,
  author={Guo, Li and Liu, Wenjian and Li, Xialin and Liu, Yixin and Jiao, Bingqi and Wang, Wei and Wang, Chengshan and Li, Fangxing},
  journal={IEEE Transactions on Smart Grid}, 
  title={Energy Management System for Stand-Alone Wind-Powered-Desalination Microgrid}, 
  year={2016},
  volume={7},
  number={2},
  pages={1079-1087},
  doi={10.1109/TSG.2014.2377374}}

@book{Mulder:96Springer,
  title={Basic Principles of Membrane Technology},
  author={Mulder, M.},
  isbn={9780792342472},
  lccn={96226059},
  year={1996},
  doi ={https://doi.org/10.1007/978-94-009-1766-8},
  publisher={Springer Netherlands}
}

@INPROCEEDINGS{Gu&Sioshansi:25PESGM,  author={Gu, Weiying and Sioshansi, Ramteen},  booktitle={2025 IEEE Power   Energy Society General Meeting (PESGM)},   title={Enhanced Convex Relaxation Approach for
Co-operated Power and Water Systems},   year={2025},  volume={},  number={},  pages={1-5}, 
  doi={10.1109/PESGM52009.2025.11225819}}

@ARTICLE{Zamzam&Emiliano&Zhao&Taylor&Sidiropoulos:19TCNS,
  author={Zamzam, Ahmed S. and Dall’Anese, Emiliano and Zhao, Changhong and Taylor, Josh A. and Sidiropoulos, Nicholas D.},
  journal={IEEE Transactions on Control of Network Systems}, 
  title={Optimal Water–Power Flow-Problem: Formulation and Distributed Optimal Solution}, 
  year={2019},
  volume={6},
  number={1},
  pages={37-47},
  doi={10.1109/TCNS.2018.2792699}}

@ARTICLE{Moazeni&Khazaei&Asrari:21TSG,
  author={Moazeni, Faegheh and Khazaei, Javad and Asrari, Arash},
  journal={IEEE Transactions on Smart Grid}, 
  title={Step Towards Energy-Water Smart Microgrids; Buildings Thermal Energy and Water Demand Management Embedded in Economic Dispatch}, 
  year={2021},
  volume={12},
  number={5},
  pages={3680-3691},
  doi={10.1109/TSG.2021.3068053}}

@ARTICLE{ZhaoetalBlockchain:23TSG,
  author={Zhao, Pengfei and Li, Shuangqi and Hu, Paul Jen-Hwa and Gu, Chenghong and Lu, Shuai and Ding, Shixing and Cao, Zhidong and Xie, Da and Xiang, Yue},
  journal={IEEE Transactions on Smart Grid}, 
  title={Blockchain-Based Water-Energy Transactive Management With Spatial-Temporal Uncertainties}, 
  year={2023},
  volume={14},
  number={4},
  pages={2903-2920},
  doi={10.1109/TSG.2022.3230693}}

@ARTICLE{CaoetalWatershed:23TSG,
  author={Cao, Yingping and Zhou, Bin and Chung, Chi Yung and Shuai, Zhikang and Hua, Zhihao and Sun, Yuexin},
  journal={IEEE Transactions on Smart Grid}, 
  title={Dynamic Modelling and Mutual Coordination of Electricity and Watershed Networks for Spatio-Temporal Operational Flexibility Enhancement Under Rainy Climates}, 
  year={2023},
  volume={14},
  number={5},
  pages={3450-3464},
  doi={10.1109/TSG.2022.3223877}}

@article{Gleick:94ARRR,
   author = "Gleick, P H",
   title = "Water and Energy", 
   journal= "Annual Review of Environment and Resources",
   year = "1994",
   volume = "19",
   number = "Volume 19, 1994",
   pages = "267-299",
   doi = "https://doi.org/10.1146/annurev.eg.19.110194.001411",
   publisher = "Annual Reviews",
   issn = "1545-2050",
   type = "Journal Article",
  }

@ARTICLE{Mohammadi&Ardakani&Abdullah&Heydt&Thomas:19TPS,
  author={Mohammadi, Farshad and Sahraei-Ardakani, Mostafa and Al-Abdullah, Yousef M. and Heydt, Gerald Thomas},
  journal={IEEE Transactions on Power Systems}, 
  title={Coordinated Scheduling of Power Generation and Water Desalination Units}, 
  year={2019},
  volume={34},
  number={5},
  pages={3657-3666},
  doi={10.1109/TPWRS.2019.2901807}}

@ARTICLE{Koraki&Strunz:18TPS,
  author={Koraki, Despina and Strunz, Kai},
  journal={IEEE Transactions on Power Systems}, 
  title={Wind and Solar Power Integration in Electricity Markets and Distribution Networks Through Service-Centric Virtual Power Plants}, 
  year={2018},
  volume={33},
  number={1},
  pages={473-485},
  doi={10.1109/TPWRS.2017.2710481}}

@ARTICLE{Jiang&Xue&Geng:13TPS,
  author={Jiang, Quanyuan and Xue, Meidong and Geng, Guangchao},
  journal={IEEE Transactions on Power Systems}, 
  title={Energy Management of Microgrid in Grid-Connected and Stand-Alone Modes}, 
  year={2013},
  volume={28},
  number={3},
  pages={3380-3389},
  doi={10.1109/TPWRS.2013.2244104}}

@article{Eke&Yusuf&Giwa&Sodiq:20Desalination,
title = {The global status of desalination: An assessment of current desalination technologies, plants and capacity},
journal = {Desalination},
volume = {495},
pages = {114633},
year = {2020},
issn = {0011-9164},
doi = {https://doi.org/10.1016/j.desal.2020.114633},
author = {Joyner Eke and Ahmed Yusuf and Adewale Giwa and Ahmed Sodiq}
}

@article{Siddiqi&Anadon:11AE,
title = {The water–energy nexus in Middle East and North Africa},
journal = {Energy Policy},
volume = {39},
number = {8},
pages = {4529-4540},
year = {2011},
issn = {0301-4215},
doi = {https://doi.org/10.1016/j.enpol.2011.04.023},
author = {Afreen Siddiqi and Laura Diaz Anadon}
}

@article{Pfenninger2016,
author = {Pfenninger, Stefan and Staffell, Iain},
doi = {10.1016/J.ENERGY.2016.08.060},
issn = {03605442},
journal = {Energy},
month = {nov},
pages = {1251--1265},
publisher = {Elsevier Ltd},
title = {{Long-term patterns of European PV output using 30 years of validated hourly reanalysis and satellite data}},
volume = {114},
year = {2016}
}

@book{Wood&Wollenberg&Sheble:13Wiley,
  author = {Wood, Allen J. and Wollenberg, Bruce F. and Shebl\'{e}, Gerald B.},
  title = {Power Generation, Operation, and Control},
  edition = {3rd}, publisher = {Wiley}, year = {2013},
  isbn= {9780471790556}
  }

@article{Zhu&Christofides&Cohen:09ACSAECR,
author = {Zhu, Aihua and Christofides, Panagiotis D and Cohen, Yoram},
doi = {10.1021/ie800735q},
issn = {0888-5885},
journal = {Industrial \& Engineering Chemistry Research},
month = {jul},
number = {13},
pages = {6010--6021},
publisher = {American Chemical Society},
title = {{Effect of Thermodynamic Restriction on Energy Cost Optimization of RO Membrane Water Desalination}},
volume = {48},
year = {2009}
}

@ARTICLE{Li&Yu&Sumaiti&Turitsyn:19TCNS,
  author={Li, Qifeng and Yu, Suhyoun and Al-Sumaiti, Ameena S. and Turitsyn, Konstantin},
  journal={IEEE Transactions on Control of Network Systems}, 
  title={Micro Water–Energy Nexus: Optimal Demand-Side Management and Quasi-Convex Hull Relaxation}, 
  year={2019},
  volume={6},
  number={4},
  pages={1313-1322},
  doi={10.1109/TCNS.2018.2889001}}
}

\appendices
%\label{app:proofs}
\section{WDP Dispatch under Renewable Uncertainty}
\label{app:MPC}

Theorem~\ref{thm:optimal} characterizes the optimal dispatch when the aggregate renewable $s=\mathbf{1}^\top\mathbf{g}$ is known. In real-time operation, the renewable generation $g_t$ is revealed sequentially over the billing period, and the period-$t$ setpoints must be dispatched before $g_t,\ldots,g_T$ are observed. The exact dispatch is then the solution of a multistage stochastic program,
\begin{equation}\label{eq:multistage}
\max_{\{\bm{x}_t = \bm{\Lambda}_t(\mathcal{F}_t)\}_{t\in\Tc}}\ \mbbE_{\mathbf{g}}\!\left[\Pi(\cdot\,;\mathbf{g})\right],
\end{equation}
where $\bm{x}_t:=(w_t^h,w_t^r,q_t^h,q_t^r,p_t^h)$ and each setpoint policy $\bm{\Lambda}_t$ is adapted to the information $\mathcal{F}_t:=\sigma(g_1,\ldots,g_{t-1})$ available when it is exercised. Problem~\eqref{eq:multistage} is, in general, intractable. Leveraging the near closed-form one-shot policy of Theorem~\ref{thm:optimal}, we adopt, in Algorithm~\ref{alg:MPC}, a receding-horizon model-predictive policy whose per-step optimization is solved exactly by a shifted four-threshold rule.

\subsubsection{Information structure and per-step problem}
\label{app:MPC-setup}
Fix a decision period $k\in\Tc$. The WDP has observed $g_1,\ldots,g_{k-1}$, has exercised the setpoints $\{w_t^{h},w_t^{r}\}_{t<k}$, and forms forecasts $\widehat{g}_k,\ldots,\widehat{g}_T$ of the future renewables. Let
\begin{equation}\label{eq:state}
\widetilde{s}_k:=\!\sum_{t=1}^{k-1} g_t,\quad
\widehat{S}_k:=\widetilde{s}_k+\!\sum_{t=k}^{T}\widehat{g}_t, \quad
\widetilde{Q}_k:=\!\sum_{t=1}^{k-1}\big(q_t^{r}-q_t^{h}\big),
\end{equation}
denote, respectively, the {\em realized} aggregate renewables, the total projected renewable over the billing period, and the {\em exercised} controllable net consumption (with $q_t^{h}=w_t^{h}/\eta_t^h$ and $q_t^{r}=(f^r)^{-1}(w_t^{r})$). Because the payment~\eqref{eq:ElectricPayment} nets over $\Bc$, the past influences the remaining decision only through the scalars in~\eqref{eq:state}. Substituting the forecasts for the unrealized renewables, the WDP solves at period $k$
\begin{equation}\label{eq:MPC}
\begin{aligned}
\Qc^{\mathsf{MPC}}_k:\quad
\max_{\{w_t^{h},w_t^{r}\}_{t\ge k}}
\;&
\pi^w\sum_{t=k}^{T}\big(w_t^{h}+w_t^{r}\big)
-\sum_{t=k}^{T}C_t^h(p_t^h)
\\
&
-P\!\Big(
\widetilde{Q}_k
+\sum_{t=k}^{T}\big(q_t^{r}-q_t^{h}\big)
-\widehat{S}_k
\Big)
\\[1mm]
\text{s.t.}\quad & \forall t\ge k\\&
\underline{w}^h\le w_t^{h}\le \overline{w}^h,\quad
\underline{w}^r\le w_t^{r}\le \overline{w}^r,
\\
&
w_t^{r}=f^r(q_t^{r}),\quad
w_t^{h}=\eta_t^h q_t^{h}=f_t^h(p_t^h).
\end{aligned}
\end{equation}
At each period only the period-$k$ setpoints are exercised; the WDP then observes $g_k$, updates~\eqref{eq:state}, and re-solves $\Qc^{\mathsf{MPC}}_{k+1}$.

\subsubsection{Shifted threshold policy}
\label{app:MPC-policy}
We define the \emph{remaining-horizon aggregate operator} $\widetilde{\Qc}_k:\mbbR\times\mbbR\to\mbbR$, as
\begin{equation}\label{eq:Pk}
\widetilde{\Qc}_k(\mu,q^r):=\widetilde{Q}_k+\sum_{t=k}^{T}\Big(q^r-\frac{\widetilde{w}_t^h(\mu)}{\eta_t^h}\Big),
\end{equation}
with $\widetilde{w}_t^h(\cdot)$ as in~\eqref{eq:wh_stationarity}, and the period-$k$ thresholds
\begin{equation}\label{eq:GammaK}
\begin{aligned}
\Gamma_{\mathsf{IM}}^{k}&:=\widetilde{\Qc}_k(\pi^+,\underline{q}^r), &
\Gamma_{\mathsf{NZ}_1}^{k}&:=\widetilde{\Qc}_k(\delta_{\mathsf{NZ}_1},\underline{q}^r),\\
\Gamma_{\mathsf{EX}}^{k}&:=\widetilde{\Qc}_k(\pi^-,\overline{q}^r), &
\Gamma_{\mathsf{NZ}_2}^{k}&:=\widetilde{\Qc}_k(\delta_{\mathsf{NZ}_2},\overline{q}^r).
\end{aligned}
\end{equation}

\begin{proposition}[Receding-horizon WDP dispatch]\label{prop:MPC}
Assume $\pi^+\!\ge\!\delta_{\mathsf{NZ}_1}\!\ge\!\delta_{\mathsf{NZ}_2}\!\ge\!\pi^-$ and the sizing condition $T(\underline{w}^h+\underline{w}^r)\ge\Wc$. For every period $k\in\Tc$, the maximizer of $\Qc^{\mathsf{MPC}}_k$ satisfies, for all $t\ge k$,
\begin{equation*}
w_t^{h\ast}=\widetilde{w}_t^h(\mu^k),\qquad 
q_t^{r\ast}=\big[\phi^r(\mu^k/\pi^w)\big]_{[\underline{q}^r,\overline{q}^r]},
\end{equation*}
where the dual $\mu^k\in[\pi^-,\pi^+]$ and the RODP setpoint obey the table in Theorem~\ref{thm:optimal} with the aggregate $s$ replaced by the projection $\widehat{S}_k$ and the thresholds $(\Gamma_{\mathsf{IM}},\Gamma_{\mathsf{NZ}_1},\Gamma_{\mathsf{NZ}_2},\Gamma_{\mathsf{EX}})$ replaced by $(\Gamma_{\mathsf{IM}}^{k},\Gamma_{\mathsf{NZ}_1}^{k},\Gamma_{\mathsf{NZ}_2}^{k},\Gamma_{\mathsf{EX}}^{k})$ of~\eqref{eq:GammaK}. In particular, $\mu^k=\pi^+$ if $\widehat{S}_k<\Gamma_{\mathsf{IM}}^{k}$, $\mu^k=\pi^-$ if $\widehat{S}_k>\Gamma_{\mathsf{EX}}^{k}$, and otherwise $\mu^k$ is the unique root in $[\pi^-,\pi^+]$ of $\widetilde{\Qc}_k(\mu,q_t^{r\ast}(\mu))=\widehat{S}_k$.
\end{proposition}

\begin{proof}
See Appendix~\ref{app:proofs}.
\end{proof}

\begin{algorithm}[t]
\caption{Receding-horizon WDP dispatch}
\label{alg:MPC}
\begin{algorithmic}[1]
\State \textbf{Input:} prices $(\pi^+,\pi^-,\pi^w)$, conversion functions $\{f_t^h\},f^r$, bounds, ratios $\{\eta_t^h\}$
\State \textbf{Initialize:} $\widetilde{s}_1\leftarrow 0$, $\widetilde{Q}_1\leftarrow 0$
\For{$k=1$ \textbf{to} $T$}
  \State forecast $\widehat{g}_k,\ldots,\widehat{g}_T$;\quad $\widehat{S}_k\leftarrow \widetilde{s}_k+\sum_{t=k}^{T}\widehat{g}_t$
  \State compute $\Gamma_{\mathsf{IM}}^{k},\Gamma_{\mathsf{NZ}_1}^{k},\Gamma_{\mathsf{NZ}_2}^{k},\Gamma_{\mathsf{EX}}^{k}$ from~\eqref{eq:Pk}--\eqref{eq:GammaK}
  \If{$\widehat{S}_k<\Gamma_{\mathsf{IM}}^{k}$}\ $\mu^k\leftarrow\pi^+$,\ $q^{r\ast}\leftarrow\underline{q}^r$
  \ElsIf{$\widehat{S}_k>\Gamma_{\mathsf{EX}}^{k}$}\ $\mu^k\leftarrow\pi^-$,\ $q^{r\ast}\leftarrow\overline{q}^r$
   \Else\ solve $\widetilde{\Qc}_k(\mu^k,q^{r\ast}(\mu^k))=\widehat{S}_k$ for $\mu^k\in[\pi^-,\pi^+]$, with $q^{r\ast}(\mu)=[\phi^r(\mu/\pi^w)]_{[\underline{q}^r,\overline{q}^r]}$
  \EndIf
  \State implement $w_k^{h}\leftarrow\widetilde{w}_k^h(\mu^k)$,\ \ $w_k^{r}\leftarrow f^r(q^{r\ast})$
  \State observe $g_k$;\ \ $\widetilde{s}_{k+1}\leftarrow\widetilde{s}_k+g_k$,\ \ $\widetilde{Q}_{k+1}\leftarrow\widetilde{Q}_k+\big(q^{r\ast}-w_k^{h}/\eta_k^h\big)$
\EndFor
\State \textbf{Output:} exercised dispatch $\{(w_k^{h},w_k^{r})\}_{k\in\Tc}$
\end{algorithmic}
\end{algorithm}

\subsubsection{Properties and implementation}
\label{app:MPC-impl}

If the forecasts are exact, \ie $\widehat{g}_t=g_t, \forall t$, then $\widehat{S}_k=s$ at every period, and $\mu^k=\mu^\ast(s)$ throughout. Hence, the receding-horizon policy reproduces the one-shot optimum (Theorem~\ref{thm:optimal}), so the only approximation relative to~\eqref{eq:multistage} is the certainty-equivalent use of forecasts, which is solved exactly. The suboptimality of Algorithm~\ref{alg:MPC}, therefore, is governed entirely by the renewable forecast error.
 
Note that, as shown in Algorithm \ref{alg:MPC}, the thresholds retain the offline character of~\eqref{eq:GammaDefs_new} and only the running offset $\widetilde{Q}_k$ and the horizon length differ across periods, so no re-optimization over past periods is required.

\subsubsection{Numerical illustration}
\label{app:MPC-num}
\begin{figure}[t]
    \centering
    \includegraphics[width=1.02\linewidth]{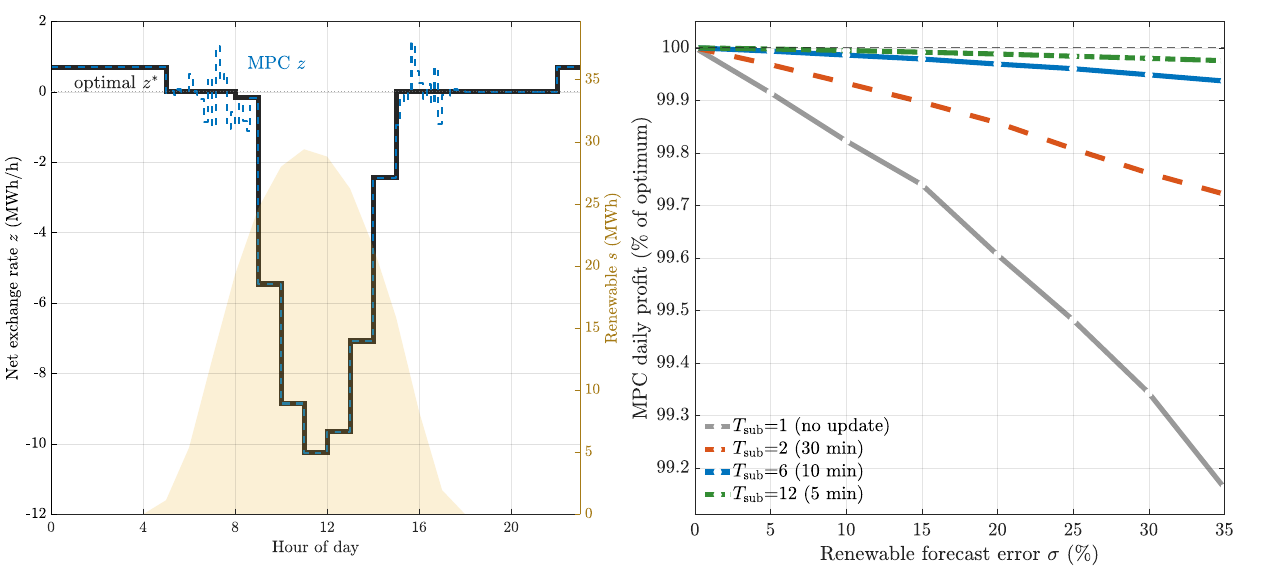}
    \vspace{-0.6cm}
    \caption{Left: net electricity exchange of the MPC dispatch (Algorithm~\ref{alg:MPC}) on a representative day at $T_{\mathrm{sub}}=6$ and $\sigma=15\%$ forecast error, versus the perfect-information optimum. Right: mean daily profit (\% of optimal) over the year versus forecast error $\sigma$, for four decision periods per netting hour.}
    \label{fig:WDPNum6}
\end{figure}

\begin{figure}[t]
    \centering
    \includegraphics[width=0.70\linewidth]{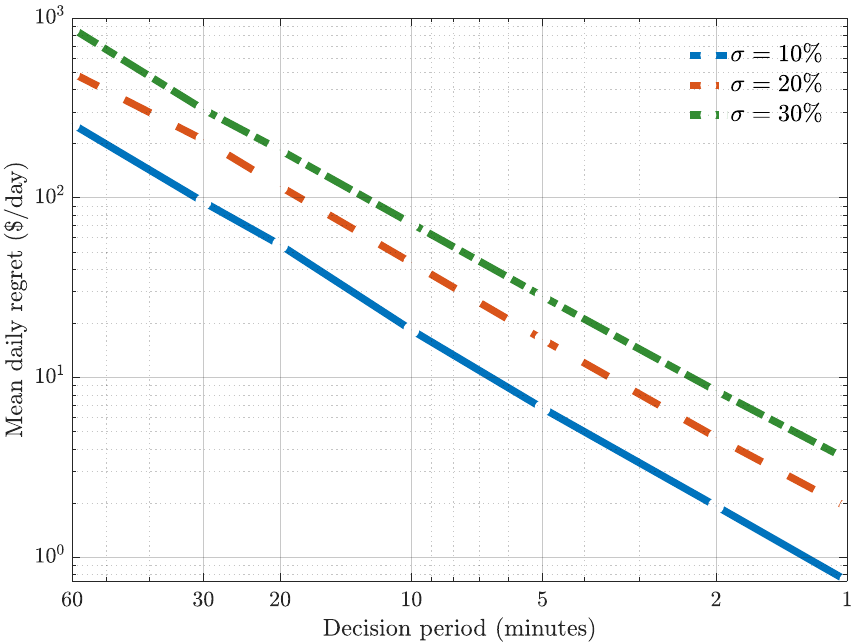}
    \vspace{-0.3cm}
    \caption{Mean daily regret of Algorithm~\ref{alg:MPC} versus the decision period under the hourly netting period for forecast errors $\sigma\in\{10\%,20\%,30\%\}$.}
    \label{fig:WDPNum7}
\end{figure}

Figures~\ref{fig:WDPNum6} and~\ref{fig:WDPNum7} apply Algorithm~\ref{alg:MPC} to the renewable data of $\S$\ref{sec:num}. The billing period is one hour with $T_{\mathrm{sub}}$ decision periods. At each step $k$, the plant observes the realized periods and forecasts each remaining period $t$ with a lead-time-scaled error $\widehat{g}_t=g_t\big(1+\sigma\sqrt{(t-k+1)/T_{\mathrm{sub}}}\,\varepsilon_{k,t}\big)$, $\varepsilon_{k,t}\sim\mathcal{N}(0,1)$; the forecast improves as the remaining window shrinks, and the current period retains error $\sigma/\sqrt{T_{\mathrm{sub}}}$. Only period $k$ is committed. By Proposition~\ref{prop:MPC}, each per-step problem is solved exactly by the one-shot threshold rule applied to the remaining horizon.

The left panel of Fig.~\ref{fig:WDPNum6} shows a representative summer day with $\sigma=15\%$ and $T_{\mathrm{sub}}=6$ (10-minute decisions). The first sub-period of each hour carries the forecast error, and the later re-solves correct it, so the committed net exchange deviates from the perfect-information trajectory only where the error pushes the projected aggregate across a threshold. The right panel shows the mean daily profit of Algorithm~\ref{alg:MPC} over the year, as a percentage of the optimal profit, as $\sigma$ increases, and $T_{\mathrm{sub}}\in\{1,2,6,12\}$. $T_{\mathrm{sub}}=1$ commits once per hour. At $\sigma=20\%$, optimality is $99.6\%$ when $T_{\mathrm{sub}}=1$, and $99.97\%$ when $T_{\mathrm{sub}}=6$.

Figure~\ref{fig:WDPNum7} shows the mean daily regret (optimal minus MPC profit) as the decision period shrinks from 60 minutes to 1 minute, for $\sigma\in\{10\%,20\%,30\%\}$. The regret falls by roughly $2.6\times$ as the decision period is halved: at $\sigma=20\%$, from about \$550/day under hourly commitment to \$40/day at 10-minute decisions and \$2/day at 1-minute decisions. Most of the gain comes from the first refinement: moving from hourly to 10-minute decisions removes over 90\% of the loss, while refining further to 1-minute decisions saves only another \$38/day.

The small losses follow from the threshold structure of Theorem~\ref{thm:optimal} rather than from forecast accuracy. Outside the net-zero region the optimal setpoints do not depend on $s$, so an error matters only if it moves the projected aggregate across a threshold. The errors are also smallest where the dispatch is most sensitive: solar-free hours are forecast exactly, and near-threshold hours have small $s$. Also, intra-hour re-solving replaces forecasts with realized energy in $\widehat{S}_k$, which is why the $T_{\mathrm{sub}}>1$ curves in Fig.~\ref{fig:WDPNum6} lie close to the optimum.

\section{Mathematical Proofs}\label{app:proofs}
\subsection{Proof of Lemma~\ref{lem:quasiconcavity}}
After substituting the equality constraints in \eqref{eq:Optimization},
namely $w_t^r = f^r(q_t^r)$, $\bm{q}^h = \bm{w}^h \oslash \bm{\eta}^h$,
and $p_t^h = (f_t^h)^{-1}(w_t^h)$ for each $t \in \Tc$
(with $\bm{\eta}^h := (\eta_t^h)_{t \in \Tc}$), the profit function in
\eqref{eq:WDPprofit} can be expressed in terms of $(\bm{w}^h, \bm{w}^r)$
alone. We analyze each term:

(i) The revenue
$R^w(\bm{w}^h, \bm{w}^r) = \pi^w \bm{1}^\top (\bm{w}^h + \bm{w}^r)$
is affine, hence concave, in $(\bm{w}^h, \bm{w}^r)$.

(ii) The electricity payment can be rewritten as
\[
    P(\bm{q}^h, \bm{q}^r; \bm{g})
    = \max\!\big(\pi^{+} z(\bm{q}^h, \bm{q}^r; \bm{g}),\;
                 \pi^{-} z(\bm{q}^h, \bm{q}^r; \bm{g})\big) + \pi^0,
\]
which, given $\pi^{+} \geq \pi^{-}$, is the pointwise maximum of two
affine functions and thus convex~\cite{Boyd&Vandenberghe:04Book}. Since
$f^r$ is strictly concave and increasing, $(f^r)^{-1}$ is strictly
convex and increasing, so
$q_t^r = (f^r)^{-1}(w_t^r)$ is convex in $w_t^r$; composition with the
affine substitutions in $\bm{w}^h$ and convex substitution in $\bm{w}^r$
preserves convexity of $P$, so $-P$ is concave in $(\bm{w}^h, \bm{w}^r)$.

(iii) The fuel cost $\sum_{t \in \Tc} C_t^h\!\left((f_t^h)^{-1}(w_t^h)\right)$ is
strictly convex in $\bm{w}^h$, since each $C_t^h(\cdot)$ is strictly
convex and non-decreasing by assumption, and $(f_t^h)^{-1}$ is strictly
convex and increasing (because $f_t^h$ is strictly concave and strictly
increasing). Hence $-\sum_{t \in \Tc} C_t^h\!\left((f_t^h)^{-1}(w_t^h)\right)$
is strictly concave in $\bm{w}^h$ and independent of $\bm{w}^r$.

Combining (i)--(iii): the sum of concave functions in $\bm{w}^r$ is
concave, and the sum of concave and strictly concave functions in
$\bm{w}^h$ is strictly concave~\cite{Boyd&Vandenberghe:04Book}.
\hfill\qedsymbol

\subsection{Proof of Theorem~\ref{thm:optimal}}
We rewrite the optimization problem~\eqref{eq:Optimization} equivalently
as
\begin{align*}
\min_{\bm{q}^r, \bm{q}^h, \bm{w}^r, \bm{w}^h} \quad
& -\pi^w \bm{1}^\top(\bm{w}^r + \bm{w}^h) + \pi^+[z]^+ - \pi^-[z]^- \\
& + \sum_{t \in \Tc} C_t^h(p_t^h) \\
\text{s.t.} \quad
& z = \bm{1}^\top(\bm{q}^r - \bm{q}^h - \bm{g}), \\
& \underline{q}^r \leq q_t^r \leq \overline{q}^r, \quad \forall\, t \in \Tc, \\
& \underline{w}^h \leq \eta_t^h q_t^h \leq \overline{w}^h, \quad \forall\, t \in \Tc,
\end{align*}
with $w_t^r = f^r(q_t^r)$, $w_t^h = \eta_t^h q_t^h$, and
$p_t^h = (f_t^h)^{-1}(\eta_t^h q_t^h)$ for all $t$.

We partition the problem into three convex and differentiable
sub-problems based on the sign of the aggregate $z$:
$\Pc^{\sf{IM}}: z>0,\;\Pc^{\sf{EX}}: z<0,\;\Pc^{\sf{NZ}}: z=0$.
Slater's condition holds for each, hence the KKT conditions are
necessary and sufficient.

\subsubsection*{Solution to $\Pc^{\sf{IM}}$}
Under $\Pc^{\sf{IM}}$, the problem becomes
\begin{align*}
\min_{\bm{q}^r, \bm{q}^h} \quad
& -\pi^w \sum_{t}\!\big(f^r(q_t^r) + \eta_t^h q_t^h\big)
+ \pi^+ \bm{1}^\top(\bm{q}^r - \bm{q}^h - \bm{g}) \\
& + \sum_{t} C_t^h(p_t^h(q_t^h)) \\
\text{s.t.} \quad
& \bm{1}^\top(\bm{q}^r - \bm{q}^h - \bm{g}) \geq 0, \\
& \underline{q}^r \leq q_t^r \leq \overline{q}^r, \quad \forall\, t, \\
& \underline{w}^h \leq \eta_t^h q_t^h \leq \overline{w}^h, \quad \forall\, t.
\end{align*}
The Lagrangian is
\begin{align*}
\Lc^{\sf{IM}}
&= -\pi^w \sum_{t}\!\big(f^r(q_t^r) + \eta_t^h q_t^h\big)
+ (\pi^+ - \mu^{\sf{IM}})\bm{1}^\top(\bm{q}^r - \bm{q}^h - \bm{g}) \\
&\quad + \sum_t C_t^h(p_t^h)
+ \sum_t \underline{\lambda}_{t,r}^{\sf{IM}}(\underline{q}^r - q_t^r) \\
&\quad + \sum_t \overline{\lambda}_{t,r}^{\sf{IM}}(q_t^r - \overline{q}^r)
+ \sum_t \underline{\lambda}_{t,h}^{\sf{IM}}(\underline{w}^h - \eta_t^h q_t^h) \\
&\quad + \sum_t \overline{\lambda}_{t,h}^{\sf{IM}}(\eta_t^h q_t^h - \overline{w}^h),
\end{align*}
with $\mu^{\sf{IM}}=0$ since $z\geq 0$ is enforced explicitly
(the active case $\mu>0$ is handled in $\Pc^{\sf{NZ}}$), and
all bound multipliers are non-negative.

Per-period stationarity in $q_t^r$ gives
\[
-\pi^w f^{\prime r}(q_t^r) + \pi^+ - \underline{\lambda}_{t,r}^{\sf{IM}}
+ \overline{\lambda}_{t,r}^{\sf{IM}} = 0.
\]
Since $\delta_{\sf{NZ}_1} = f^{\prime r}(\underline{q}^r)\pi^w \leq \pi^+$ by
assumption and $f^{\prime r}$ is strictly decreasing,
$\pi^w f^{\prime r}(q_t^r) \leq \delta_{\sf{NZ}_1} \leq \pi^+$ for all
$q_t^r \in [\underline{q}^r, \overline{q}^r]$, so the coefficient
$-\pi^w f^{\prime r}(q_t^r) + \pi^+ \geq 0$. This forces
$\underline{\lambda}_{t,r}^{\sf{IM}} \geq 0$ and saturates the lower
bound: $q_t^{r,\sf{IM}} = \underline{q}^r$, hence
$w_t^{r,\sf{IM}} = \underline{w}^r$ for all $t$.

Per-period stationarity in $q_t^h$ uses
$\partial p_t^h / \partial q_t^h = \eta_t^h / f_t^{\prime h}(p_t^h)$
(from differentiating $\eta_t^h q_t^h = f_t^h(p_t^h)$), giving
\begin{align*}
\frac{\partial \Lc^{\sf{IM}}}{\partial q_t^h}
&= -\eta_t^h \pi^w - \pi^+ + \frac{\eta_t^h}{f_t^{\prime h}(p_t^h)}(C_t^h)'(p_t^h) \\
&\quad - \eta_t^h \underline{\lambda}_{t,h}^{\sf{IM}}
+ \eta_t^h \overline{\lambda}_{t,h}^{\sf{IM}} = 0.
\end{align*}
Multiplying by $f_t^{\prime h}(p_t^h)/\eta_t^h$ and using
$\beta_t^h(p_t^h) = f_t^{\prime h}(p_t^h)/\eta_t^h$, the interior
stationarity condition becomes
\[
(C_t^h)'(p_t^h) = f_t^{\prime h}(p_t^h)\, \pi^w + \beta_t^h(p_t^h)\, \pi^+,
\]
which is precisely~\eqref{eq:p_implicit} evaluated at $\mu = \pi^+$.
This implicit equation admits a unique solution: the left-hand side is
strictly increasing in $p$ (since $C_t^h$ is strictly convex), while
the right-hand side $f_t^{\prime h}(p)(\pi^w + \pi^+/\eta_t^h)$ is strictly
decreasing in $p$ (since $f_t^{\prime h}$ is strictly decreasing and
$\pi^w + \pi^+/\eta_t^h > 0$). Projecting onto the TDP bounds yields
\[
w_t^{h,\sf{IM}} := \eta_t^h q_t^{h,\sf{IM}}
= \big[f_t^h(p_t^h(\pi^+))\big]_{[\underline{w}^h,\overline{w}^h]}
= \widetilde{w}_t^h(\pi^+),
\]
matching~\eqref{eq:wh_stationarity} evaluated at $\mu = \pi^+$. In the
linear case $f_t^h(p)=\alpha_t^h p$, the FOC reduces to
$(C_t^h)'(p) = \alpha_t^h \pi^w + (\alpha_t^h/\eta_t^h)\pi^+$, giving
the closed form
$w_t^{h,\sf{IM}} = \big[\alpha_t^h D_t^h(\alpha_t^h \pi^w + (\alpha_t^h/\eta_t^h)\pi^+)\big]_{[\underline{w}^h,\overline{w}^h]}$.

The IM solution is feasible only if $z^{\sf{IM}}(\bm{g}) > 0$, \ie
\[
\bm{1}^\top \bm{g} \;<\;
T\underline{q}^r - \sum_t \frac{\widetilde{w}_t^h(\pi^+)}{\eta_t^h}
\;=\; \Qc(\pi^+, \underline{q}^r)
\;=\; \Gamma_{\sf{IM}},
\]
where the last equality uses \eqref{eq:AggPowerOp} and
\eqref{eq:GammaDefs_new}.

\subsubsection*{Solution to $\Pc^{\sf{EX}}$}
By symmetry to $\Pc^{\sf{IM}}$, with $\mu^{\sf{EX}}=0$ and using the
assumption $\delta_{\sf{NZ}_2} = f^{\prime r}(\overline{q}^r)\pi^w \geq \pi^-$:
since $f^{\prime r}$ is strictly decreasing,
$\pi^w f^{\prime r}(q_t^r) \geq \delta_{\sf{NZ}_2} \geq \pi^-$ throughout, so
the stationarity coefficient $-\pi^w f^{\prime r}(q_t^r)+\pi^- \leq 0$,
saturating the upper bound: $w_t^{r,\sf{EX}} = \overline{w}^r$ for all
$t$. The TDP setpoint, derived from~\eqref{eq:p_implicit} at
$\mu = \pi^-$, is $w_t^{h,\sf{EX}} = \widetilde{w}_t^h(\pi^-)$.
Feasibility requires $z^{\sf{EX}}(\bm{g}) < 0$, \ie
\[
\bm{1}^\top \bm{g} \;>\;
T\overline{q}^r - \sum_t \frac{\widetilde{w}_t^h(\pi^-)}{\eta_t^h}
\;=\; \Qc(\pi^-, \overline{q}^r)
\;=\; \Gamma_{\sf{EX}}.
\]

\subsubsection*{Solution to $\Pc^{\sf{NZ}}$}
Under $\Pc^{\sf{NZ}}$, the problem becomes
\begin{align*}
\min_{\bm{q}^r, \bm{q}^h} \quad
& -\pi^w \sum_{t}\!\big(f^r(q_t^r) + \eta_t^h q_t^h\big) + \sum_t C_t^h(p_t^h) \\
\text{s.t.} \quad
& \bm{1}^\top(\bm{q}^r - \bm{q}^h - \bm{g}) = 0, \\
& \underline{q}^r \leq q_t^r \leq \overline{q}^r, \quad \forall\, t, \\
& \underline{w}^h \leq \eta_t^h q_t^h \leq \overline{w}^h, \quad \forall\, t.
\end{align*}
Let $\mu$ denote the single dual variable associated with the net consumption equality, and let
$\underline{\lambda}_{t,r}, \overline{\lambda}_{t,r},
\underline{\lambda}_{t,h}, \overline{\lambda}_{t,h} \geq 0$ be the bound
multipliers (region superscripts removed for brevity). Per-period stationarity:
\begin{align}
\partial_{q_t^r} \Lc^{\sf{NZ}}
&= -\pi^w f^{\prime r}(q_t^r) + \mu
- \underline{\lambda}_{t,r}
+ \overline{\lambda}_{t,r} = 0, \label{eq:st-r-NZ}\\
\partial_{q_t^h} \Lc^{\sf{NZ}}
&= -\eta_t^h \pi^w - \mu
+ \frac{\eta_t^h}{f_t^{\prime h}(p_t^h)}(C_t^h)'(p_t^h) \notag \\
&\quad - \eta_t^h \underline{\lambda}_{t,h}
+ \eta_t^h \overline{\lambda}_{t,h} = 0. \label{eq:st-h-NZ}
\end{align}

\noindent Note that the TDP per-period stationarity~\eqref{eq:st-h-NZ} depends
on the dual $\mu$ alone and does not involve the RODP multipliers
$\underline{\lambda}_{t,r}, \overline{\lambda}_{t,r}$. Multiplying
through by $f_t^{\prime h}(p_t^h)/\eta_t^h$ recovers the implicit
equation~\eqref{eq:p_implicit}, and projecting onto the TDP bounds
therefore yields, in every NZ sub-segment,
\begin{equation}
w_t^{h\ast} = \widetilde{w}_t^h(\mu),
\label{eq:NZ-h-uniform}
\end{equation}
as in~\eqref{eq:wh_stationarity}. Moreover, $\widetilde{w}_t^h$ is
non-decreasing in $\mu$: by the same monotonicity argument as in the IM
solution, increasing $\mu$ shifts the right-hand side of
\eqref{eq:p_implicit} up at every $p$, so $p_t^h(\mu)$ is non-decreasing
in $\mu$, and hence so is $f_t^h(p_t^h(\mu))$. What differs across
sub-segments is the value of the dual and the RODP setpoint, jointly
pinned by the RODP stationarity~\eqref{eq:st-r-NZ} and the net zero
consumption. We denote the dual by $\mu^{\sf{NZ}}$, $\mu^{\sf{NZ}_1}$,
and $\mu^{\sf{NZ}_2}$ in the NZ-interior, NZ-lower, and NZ-upper
sub-segments, respectively.

\paragraph{Interior NZ solution.}
Suppose all RODP bound multipliers vanish. Then~\eqref{eq:st-r-NZ} gives
$f^{\prime r}(q_t^r) = \mu^{\sf{NZ}}/\pi^w$, hence
$q_t^{r\ast} = \phi^r\!\big(\mu^{\sf{NZ}}/\pi^w\big)$; uniform
across $t$ since the right-hand side is. The net zero equality
$T q_t^{r\ast} - \sum_t q_t^{h\ast} = \bm{1}^\top \bm{g}$ together
with~\eqref{eq:NZ-h-uniform} pins $\mu^{\sf{NZ}}$ via
\begin{equation}\label{eq:muNZ-interior-proof}
T\phi^r\!\big(\mu/\pi^w\big)
- \sum_t \frac{\widetilde{w}_t^h(\mu)}{\eta_t^h}
= \bm{1}^\top \bm{g},
\end{equation}
matching the implicit equation defining $\mu^{\sf{NZ}}$ in
Theorem~\ref{thm:optimal}. The left-hand side is strictly decreasing in
$\mu$ (since $\phi^r$ is strictly decreasing and each
$\widetilde{w}_t^h$ is non-decreasing in $\mu$), hence
$\mu^{\sf{NZ}}(\bm{1}^\top \bm{g})$ is unique. Setting
$w_t^{r\ast} = f^r(q_t^{r\ast})
            = f^r\!\big(\phi^r(\mu^{\sf{NZ}}/\pi^w)\big)$
recovers the NZ-interior RODP setpoint in Theorem~\ref{thm:optimal}.
The interior NZ solution requires
$\underline{q}^r \leq q_t^{r\ast} \leq \overline{q}^r$, equivalently
$\mu^{\sf{NZ}} \in [\delta_{\sf{NZ}_2}, \delta_{\sf{NZ}_1}]$, which by
monotonicity of \eqref{eq:muNZ-interior-proof} corresponds to
$\Gamma_{\sf{NZ}_1} \leq \bm{1}^\top \bm{g} \leq \Gamma_{\sf{NZ}_2}$,
with $\Gamma_{\sf{NZ}_1}, \Gamma_{\sf{NZ}_2}$ as
in~\eqref{eq:GammaDefs_new}.

\paragraph{Boundary NZ-lower solution.}
For $\bm{1}^\top \bm{g} \in [\Gamma_{\sf{IM}}, \Gamma_{\sf{NZ}_1})$, the
RODP lower bound is active: $q_t^{r\ast} = \underline{q}^r$ (so
$w_t^{r\ast} = \underline{w}^r$) at every $t$, with
$\underline{\lambda}_{t,r} = \mu^{\sf{NZ}_1} - \delta_{\sf{NZ}_1} \geq 0$
from~\eqref{eq:st-r-NZ}. Since~\eqref{eq:st-h-NZ} involves the dual
only, the TDP setpoints retain the form~\eqref{eq:NZ-h-uniform}:
$w_t^{h\ast} = \widetilde{w}_t^h(\mu^{\sf{NZ}_1})$ for every $t$. The
dual $\mu^{\sf{NZ}_1}$ is pinned by the net zero balance with the RODP
fixed at $\underline{q}^r$:
\begin{equation}\label{eq:muNZ1-implicit}
T\underline{q}^r - \sum_t \frac{\widetilde{w}_t^h(\mu)}{\eta_t^h}
= \Qc(\mu,\underline{q}^r)
= \bm{1}^\top \bm{g},
\quad \mu^{\sf{NZ}_1} \in [\delta_{\sf{NZ}_1}, \pi^+],
\end{equation}
matching the implicit equation defining $\mu^{\sf{NZ}_1}$ in
Theorem~\ref{thm:optimal}. By monotonicity of $\Qc(\cdot,\underline{q}^r)$
in $\mu$, the solution is unique; it traces $\pi^+$ at
$\bm{1}^\top \bm{g} = \Gamma_{\sf{IM}}$ and $\delta_{\sf{NZ}_1}$ at
$\bm{1}^\top \bm{g} = \Gamma_{\sf{NZ}_1}$, ensuring continuity with the
IM regime and the NZ-interior. Notice that we do \emph{not} need to
enforce per-period balance $q_t^h = \underline{q}^r - g_t$; the
net zero balance combined with the cost-minimizing TDP
allocation~\eqref{eq:NZ-h-uniform} suffices.

\paragraph{Boundary NZ-upper solution.}
Symmetrically, for
$\bm{1}^\top \bm{g} \in (\Gamma_{\sf{NZ}_2}, \Gamma_{\sf{EX}}]$, the RODP
upper bound is active: $w_t^{r\ast} = \overline{w}^r$ and
$w_t^{h\ast} = \widetilde{w}_t^h(\mu^{\sf{NZ}_2})$, with
$\overline{\lambda}_{t,r} = \delta_{\sf{NZ}_2} - \mu^{\sf{NZ}_2} \geq 0$.
The dual $\mu^{\sf{NZ}_2} \in [\pi^-, \delta_{\sf{NZ}_2}]$ is pinned by
\begin{equation}\label{eq:muNZ2-implicit}
T\overline{q}^r - \sum_t \frac{\widetilde{w}_t^h(\mu)}{\eta_t^h}
= \Qc(\mu,\overline{q}^r)
= \bm{1}^\top \bm{g},
\end{equation}
matching the implicit equation defining $\mu^{\sf{NZ}_2}$ in
Theorem~\ref{thm:optimal}.

\paragraph{Selection of the optimal regime.}
By convexity and Slater's condition, each region's KKT point is globally
optimal within its region. Combining the IM, EX, and NZ feasibility
conditions:
\[
\bm{1}^\top \bm{g} \mapsto
\begin{cases}
\mbox{IM}          & \bm{1}^\top \bm{g} < \Gamma_{\sf{IM}}, \\
\mbox{NZ-lower}    & \Gamma_{\sf{IM}} \leq \bm{1}^\top \bm{g} < \Gamma_{\sf{NZ}_1}, \\
\mbox{NZ-interior} & \Gamma_{\sf{NZ}_1} \leq \bm{1}^\top \bm{g} \leq \Gamma_{\sf{NZ}_2}, \\
\mbox{NZ-upper}    & \Gamma_{\sf{NZ}_2} < \bm{1}^\top \bm{g} \leq \Gamma_{\sf{EX}}, \\
\mbox{EX}          & \bm{1}^\top \bm{g} > \Gamma_{\sf{EX}},
\end{cases}
\]
yielding the dispatch tabulated in Theorem~\ref{thm:optimal}.

\paragraph{Remark (equality cases).}
At $\delta_{\sf{NZ}_1} = \pi^+$, the IM stationarity in $q_t^r$ is flat
at $\underline{q}^r$, so any $w_t^r$ in a neighborhood is optimal at
that price; the limit under an arbitrarily small increase of $\pi^+$
selects $w_t^r = \underline{w}^r$, and a small decrease selects
$w_t^r > \underline{w}^r$. A symmetric statement holds at
$\delta_{\sf{NZ}_2} = \pi^-$.
\hfill\qedsymbol

\subsection{Proof of Corollary~\ref{prop:Optimalz}}
By definition, the aggregate net consumption is
\[
z^\ast(\bm{g}) = \bm{1}^\top(\bm{q}^{r\ast}-\bm{q}^{h\ast}-\bm{g})
= \sum_t (f^r)^{-1}(w_t^{r\ast}) - \sum_t \frac{w_t^{h\ast}}{\eta_t^h} - \bm{1}^\top \bm{g}.
\]
Substituting the optimal dispatch tabulated in
Theorem~\ref{thm:optimal}:
\begin{itemize}[leftmargin=*]
\item For $\bm{1}^\top \bm{g} < \Gamma_{\sf{IM}}$ (IM regime):
$w_t^{r\ast} = \underline{w}^r$ (so $(f^r)^{-1}(w_t^{r\ast}) = \underline{q}^r$)
and $w_t^{h\ast} = \widetilde{w}_t^h(\pi^+)$, hence
$z^\ast(\bm{g}) = T\underline{q}^r - \sum_t \widetilde{w}_t^h(\pi^+)/\eta_t^h - \bm{1}^\top \bm{g}
= \Gamma_{\sf{IM}} - \bm{1}^\top \bm{g} > 0$.

\item For $\bm{1}^\top \bm{g} \in [\Gamma_{\sf{IM}}, \Gamma_{\sf{EX}}]$
(NZ regimes): the net zero balance
$\bm{1}^\top \bm{q}^{r\ast} = \bm{1}^\top \bm{q}^{h\ast} + \bm{1}^\top \bm{g}$
holds by construction, so $z^\ast(\bm{g}) = 0$.

\item For $\bm{1}^\top \bm{g} > \Gamma_{\sf{EX}}$ (EX regime):
$w_t^{r\ast} = \overline{w}^r$ and $w_t^{h\ast} = \widetilde{w}_t^h(\pi^-)$,
hence $z^\ast(\bm{g}) = \Gamma_{\sf{EX}} - \bm{1}^\top \bm{g} < 0$.
\end{itemize}
The sign function in~\eqref{eq:Optz} follows. \hfill\qedsymbol

\subsection{Proof of Proposition~\ref{prop:SpecialCases}}
In every sub-problem $\sigma \in \{\sf{IM}, \sf{NZ}, \sf{EX}\}$ the per-period RODP
stationarity reads
\begin{equation}\label{eq:rodp-stat-prop}
\begin{aligned}
-\pi^w f^{\prime r}(q_t^r) + \delta_\sigma - \underline{\lambda}_{t,r}
+ \overline{\lambda}_{t,r} = 0,\\[-1pt]
\delta_{\sf{IM}} = \pi^+,\quad
\delta_{\sf{EX}} = \pi^-,\quad
\delta_{\sf{NZ}} = \mu \in [\pi^-, \pi^+].
\end{aligned}
\end{equation}
with non-negative bound multipliers, and the TDP setpoint is
$\widetilde{w}_t^h(\delta_\sigma)$ throughout, as established in the proof
of Theorem~\ref{thm:optimal}.

\paragraph{Case 1: $\delta_{\sf{NZ}_2} > \pi^+$ (RODP saturated at maximum).}
By definition $\delta_{\sf{NZ}_2} = f^{\prime r}(\overline{q}^r)\pi^w$. The assumption $f^{\prime r}(\overline{q}^r)\pi^w > \pi^+ \geq \pi^-$, combined with the strict monotonicity of $f^{\prime r}$, gives
\[
\pi^w f^{\prime r}(q_t^r) \;\geq\; \pi^w f^{\prime r}(\overline{q}^r) \;=\; \delta_{\sf{NZ}_2} \;>\; \pi^+,
\qquad \forall\, q_t^r \in [\underline{q}^r, \overline{q}^r].
\]
In each sub-problem $\sigma$, since
$\delta_\sigma \leq \pi^+ < \pi^w f^{\prime r}(q_t^r)$ throughout the operating
range, the coefficient of~\eqref{eq:rodp-stat-prop},
$-\pi^w f^{\prime r}(q_t^r) + \delta_\sigma < 0$ everywhere. With non-negative
bound multipliers this forces $\overline{\lambda}_{t,r} > 0$, saturating
the upper bound: $q_t^{r\ast} = \overline{q}^r$, hence
$w_t^{r\ast} = \overline{w}^r$ for all $t$.

\paragraph{Case 2: $\delta_{\sf{NZ}_1} < \pi^-$ (RODP saturated at minimum).}
Symmetrically, $f^{\prime r}(\underline{q}^r)\pi^w < \pi^- \leq \pi^+$ and the
strict monotonicity of $f^{\prime r}$ give
\[
\pi^w f^{\prime r}(q_t^r) \;\leq\; \pi^w f^{\prime r}(\underline{q}^r) \;=\; \delta_{\sf{NZ}_1} \;<\; \pi^-,
\qquad \forall\, q_t^r \in [\underline{q}^r, \overline{q}^r].
\]
In each sub-problem, $\delta_\sigma \geq \pi^- > \pi^w f^{\prime r}(q_t^r)$,
so the coefficient $-\pi^w f^{\prime r}(q_t^r) + \delta_\sigma > 0$, forcing
$\underline{\lambda}_{t,r} > 0$ and saturating the lower bound:
$q_t^{r\ast} = \underline{q}^r$, hence
$w_t^{r\ast} = \underline{w}^r$ for all $t$.

Together, Cases~1 and~2 establish~\eqref{eq:OptWrSpecial}.

\paragraph{TDP dispatch.}
With $\bm{w}^{r\ast}$ fixed, the residual TDP problem is to choose
$\bm{w}^h$ to minimize
$-\pi^w \bm{1}^\top \bm{w}^h + \pi^+[\tilde{z}]^+ - \pi^-[\tilde{z}]^- + \sum_t C_t^h$,
subject to the TDP bounds, where
$\tilde{z} = T(f^r)^{-1}(w^{r\ast}) - \sum_t q_t^h - \bm{1}^\top \bm{g}$.
The same three-region (IM/NZ/EX) analysis as in
Theorem~\ref{thm:optimal} applies, now with no degrees of freedom in
the RODP. In the NZ region, per-period stationarity in $q_t^h$ yields
$w_t^{h\ast} = \widetilde{w}_t^h(\mu^{\sf{s}})$ as
in~\eqref{eq:NZ-h-uniform}, with the dual
$\mu^{\sf{s}} \in [\pi^-, \pi^+]$ pinned by the net zero balance
\[
T(f^r)^{-1}(w^{r\ast}) - \sum_t \frac{\widetilde{w}_t^h(\mu)}{\eta_t^h}
\;=\; \Qc\!\big(\mu,\,(f^r)^{-1}(w^{r\ast})\big)
\;=\; \bm{1}^\top \bm{g}.
\]
The IM and EX setpoints reduce to $\widetilde{w}_t^h(\pi^+)$ and
$\widetilde{w}_t^h(\pi^-)$ respectively. The corresponding thresholds
are
\[
\Gamma_{\sf{IM}}^{\sf{s}} = \Qc\!\big(\pi^+,\,(f^r)^{-1}(w^{r\ast})\big),
\quad
\Gamma_{\sf{EX}}^{\sf{s}} = \Qc\!\big(\pi^-,\,(f^r)^{-1}(w^{r\ast})\big),
\]
matching the thresholds in Proposition~\ref{prop:SpecialCases}. The
resulting dispatch is~\eqref{eq:OptWhSpecial}.

When $\pi^+ = \pi^-$, the coefficients on $q_t^h$ in IM and EX coincide,
so $\widetilde{w}_t^h(\pi^+) = \widetilde{w}_t^h(\pi^-)$, and the three
regions collapse:
$w_t^{h\ast} = \widetilde{w}_t^h(\pi^+) = \widetilde{w}_t^h(\pi^-)$ for
all $\bm{g}$.
\hfill\qedsymbol

\subsection{Extended Proposition~\ref{prop:SpecialCases}: Interior RODP}
\label{app:SpecialCasesExt}

Proposition~\ref{prop:SpecialCases} treats the case where the price interval lies entirely outside the band $(\delta_{\mathsf{NZ}_2}, \delta_{\mathsf{NZ}_1})$. The following extends it to the complementary case, where a grid price falls strictly inside this band and the RODP is interior at that price. All such cases are governed by the \emph{clamped RODP setpoint}
\begin{equation}\label{eq:qr_hat}
    \widehat{q}^r(\mu) := \big[\phi^r(\mu/\pi^w)\big]_{[\underline{q}^r,\, \overline{q}^r]},
\end{equation}
which, since $f^{\prime r}$ is strictly decreasing, equals $\underline{q}^r$ for $\mu \geq \delta_{\mathsf{NZ}_1}$, equals $\overline{q}^r$ for $\mu \leq \delta_{\mathsf{NZ}_2}$, and is interior otherwise.

\begin{proposition}[Optimal dispatch under imbalanced tariffs: interior RODP]
\label{prop:SpecialCasesExt}
When a grid price lies inside $(\delta_{\mathsf{NZ}_2}, \delta_{\mathsf{NZ}_1})$, the RODP is interior at that boundary, giving three sub-cases:

\emph{Case A} $(\pi^-<\delta_{\mathsf{NZ}_2}\leq\pi^+<\delta_{\mathsf{NZ}_1})$: $\Gamma_{\mathsf{IM}}=\Qc(\pi^+,\widehat{q}^r(\pi^+))$; $\Gamma_{\mathsf{NZ}_2},\Gamma_{\mathsf{EX}}$ as in \eqref{eq:GammaDefs_new}.
\begin{center}\footnotesize\normalfont\renewcommand{\arraystretch}{1.4}
\begin{tabular}{@{}llll@{}}
\toprule
\em{Mode} & \em{Range of} $s$ & \em{TDP} $w_t^{h\ast}(s)$ & \em{RODP} $w_t^{r\ast}(s)$ \\ \midrule
IM & $[0, \Gamma_{\mathsf{IM}})$ & $\widetilde{w}_t^h(\pi^+)$ & $f^r(\widehat{q}^r(\pi^+))$ \\
NZ & $[\Gamma_{\mathsf{IM}}, \Gamma_{\mathsf{NZ}_2}]$ & $\widetilde{w}_t^h(\mu^{\mathsf{NZ}}(s))$ & $f^r(\phi^r(\frac{\mu^{\mathsf{NZ}}(s)}{\pi^w}))$ \\
NZ & $(\Gamma_{\mathsf{NZ}_2}, \Gamma_{\mathsf{EX}}]$ & $\widetilde{w}_t^h(\mu^{\mathsf{NZ}_2}(s))$ & $\overline{w}^r$ \\
EX & $(\Gamma_{\mathsf{EX}}, \infty)$ & $\widetilde{w}_t^h(\pi^-)$ & $\overline{w}^r$ \\ \bottomrule
\end{tabular}
\end{center}

\emph{Case B} $(\delta_{\mathsf{NZ}_2}<\pi^-\leq\delta_{\mathsf{NZ}_1}\leq\pi^+)$: $\Gamma_{\mathsf{EX}}=\Qc(\pi^-,\widehat{q}^r(\pi^-))$; $\Gamma_{\mathsf{IM}},\Gamma_{\mathsf{NZ}_1}$ as in \eqref{eq:GammaDefs_new}.
\begin{center}\footnotesize\normalfont\renewcommand{\arraystretch}{1.4}
\begin{tabular}{@{}llll@{}}
\toprule
\em{Mode} & \em{Range of} $s$ & \em{TDP} $w_t^{h\ast}(s)$ & \em{RODP} $w_t^{r\ast}(s)$ \\ \midrule
IM & $[0, \Gamma_{\mathsf{IM}})$ & $\widetilde{w}_t^h(\pi^+)$ & $\underline{w}^r$ \\
NZ & $[\Gamma_{\mathsf{IM}}, \Gamma_{\mathsf{NZ}_1})$ & $\widetilde{w}_t^h(\mu^{\mathsf{NZ}_1}(s))$ & $\underline{w}^r$ \\
NZ & $[\Gamma_{\mathsf{NZ}_1}, \Gamma_{\mathsf{EX}}]$ & $\widetilde{w}_t^h(\mu^{\mathsf{NZ}}(s))$ & $f^r(\phi^r(\frac{\mu^{\mathsf{NZ}}(s)}{\pi^w}))$ \\
EX & $(\Gamma_{\mathsf{EX}}, \infty)$ & $\widetilde{w}_t^h(\pi^-)$ & $f^r(\widehat{q}^r(\pi^-))$ \\ \bottomrule
\end{tabular}
\end{center}

\emph{Case C} $(\delta_{\mathsf{NZ}_2}<\pi^-\leq\pi^+<\delta_{\mathsf{NZ}_1})$: $\Gamma_{\mathsf{IM}}=\Qc(\pi^+,\widehat{q}^r(\pi^+))$, $\Gamma_{\mathsf{EX}}=\Qc(\pi^-,\widehat{q}^r(\pi^-))$.
\begin{center}\footnotesize\normalfont\renewcommand{\arraystretch}{1.4}
\begin{tabular}{@{}llll@{}}
\toprule
\em{Mode} & \em{Range of} $s$ & \em{TDP} $w_t^{h\ast}(s)$ & \em{RODP} $w_t^{r\ast}(s)$ \\ \midrule
IM & $[0, \Gamma_{\mathsf{IM}})$ & $\widetilde{w}_t^h(\pi^+)$ & $f^r(\widehat{q}^r(\pi^+))$ \\
NZ & $[\Gamma_{\mathsf{IM}}, \Gamma_{\mathsf{EX}}]$ & $\widetilde{w}_t^h(\mu^{\mathsf{NZ}}(s))$ & $f^r(\phi^r(\frac{\mu^{\mathsf{NZ}}(s)}{\pi^w}))$ \\
EX & $(\Gamma_{\mathsf{EX}}, \infty)$ & $\widetilde{w}_t^h(\pi^-)$ & $f^r(\widehat{q}^r(\pi^-))$ \\ \bottomrule
\end{tabular}
\end{center}
In Case~A the plant imports while still producing RO water ($f^r(\widehat{q}^r(\pi^+))>\underline{w}^r$ in the IM mode); Case~B is its export-side mirror; and Case~C, with both prices interior, reduces to a two-threshold policy in which the RODP is interior throughout.
\end{proposition}

\begin{proof}
Recall the per-period RODP stationarity condition~\eqref{eq:rodp-stat-prop} established in the proof of Proposition~\ref{prop:SpecialCases}. When neither Case~1 nor Case~2 holds, at least one grid price lies strictly inside
$(\delta_{\sf{NZ}_2}, \delta_{\sf{NZ}_1})$. Consider a
sub-problem $\sigma$ with faced price
$\delta_\sigma \in (\delta_{\sf{NZ}_2}, \delta_{\sf{NZ}_1})$. Then
$\pi^w f^{\prime r}(\overline{q}^r) = \delta_{\sf{NZ}_2} < \delta_\sigma
< \delta_{\sf{NZ}_1} = \pi^w f^{\prime r}(\underline{q}^r)$, so the coefficient
of~\eqref{eq:rodp-stat-prop}, $-\pi^w f^{\prime r}(q_t^r) + \delta_\sigma$, is
negative at $q_t^r = \underline{q}^r$ and positive at
$q_t^r = \overline{q}^r$; by strict monotonicity of $f^{\prime r}$ it vanishes
at the unique interior point
$q_t^{r\ast} = \phi^r(\delta_\sigma/\pi^w)$, with both bound
multipliers zero. Combined with Cases~1--2, the RODP setpoint at any
faced price is the clamped value
\begin{equation}\label{eq:qrhat-proof}
q_t^{r\ast}
= \big[\phi^r(\delta_\sigma/\pi^w)\big]_{[\underline{q}^r,\overline{q}^r]}
= \widehat{q}^r(\delta_\sigma),
\end{equation}
as in~\eqref{eq:qr_hat}: the lower bound is active iff
$\delta_\sigma \geq \delta_{\sf{NZ}_1}$, the upper bound iff
$\delta_\sigma \leq \delta_{\sf{NZ}_2}$, and $q_t^{r\ast}$ is interior
otherwise. Setting $\delta_\sigma = \pi^+$ (IM) and $\delta_\sigma = \pi^-$
(EX) in the feasibility conditions of Theorem~\ref{thm:optimal} fixes the
boundaries $\Gamma_{\sf{IM}} = \Qc(\pi^+, \widehat{q}^r(\pi^+))$ and
$\Gamma_{\sf{EX}} = \Qc(\pi^-, \widehat{q}^r(\pi^-))$; within NZ the
setpoint $\widehat{q}^r(\mu)$ departs a bound only as the dual $\mu$
crosses $\delta_{\sf{NZ}_1}$ or $\delta_{\sf{NZ}_2}$, \ie at
$\Gamma_{\sf{NZ}_1}$ or $\Gamma_{\sf{NZ}_2}$ of~\eqref{eq:GammaDefs_new}, and such a crossing occurs within NZ only if the corresponding
threshold price lies in $[\pi^-, \pi^+]$. Since $\mu$ ranges over
$[\pi^-, \pi^+]$, three sub-cases follow.

\paragraph{Case A: $\pi^- < \delta_{\sf{NZ}_2} \le \pi^+ < \delta_{\sf{NZ}_1}$.}
Here $\widehat{q}^r(\pi^+)$ is interior while
$\widehat{q}^r(\pi^-) = \overline{q}^r$. As $\mu \in [\pi^-, \pi^+]$ with
$\pi^+ < \delta_{\sf{NZ}_1}$, the RODP lower bound is never active and the
NZ-lower segment vanishes; as $\mu$ decreases through
$\delta_{\sf{NZ}_2} \in (\pi^-, \pi^+]$ the RODP saturates its upper
bound, at $\Gamma_{\sf{NZ}_2}$. The surviving IM, NZ-interior, NZ-upper,
and EX segments reproduce the Case~A table.

\paragraph{Case B: $\delta_{\sf{NZ}_2} < \pi^- \le \delta_{\sf{NZ}_1} \le \pi^+$.}
By the mirror argument, $\widehat{q}^r(\pi^+) = \underline{q}^r$ and
$\widehat{q}^r(\pi^-)$ is interior; the NZ-upper segment vanishes (since
$\mu \geq \pi^- > \delta_{\sf{NZ}_2}$) and the RODP leaves its lower bound
at $\Gamma_{\sf{NZ}_1}$. The surviving IM, NZ-lower, NZ-interior, and EX
segments reproduce the Case~B table.

\paragraph{Case C: $\delta_{\sf{NZ}_2} < \pi^- \le \pi^+ < \delta_{\sf{NZ}_1}$.}
Both faced prices are interior, so $[\pi^-, \pi^+] \subset
(\delta_{\sf{NZ}_2}, \delta_{\sf{NZ}_1})$ and neither RODP bound is active
for any $\mu$; both $\Gamma_{\sf{NZ}_1}$ and $\Gamma_{\sf{NZ}_2}$ lie
outside $[\Gamma_{\sf{IM}}, \Gamma_{\sf{EX}}]$, leaving only the IM,
NZ-interior, and EX segments (Case~C table).

Continuity across the surviving segment boundaries follows, as in the
proof of Theorem~\ref{thm:optimal}, from the shared dual value at each
threshold.
\hfill\qedsymbol
\end{proof}

\subsection{Proof of Corollary~\ref{corol:RODPTDPonly}}
The proof follows directly from Theorem~\ref{thm:optimal}.

\paragraph{RODP-only case.}
All TDP terms vanish ($w_t^h = 0$ for all $t$), so
$\widetilde{w}_t^h(\cdot) \equiv 0$ and
from~\eqref{eq:GammaDefs_new} the four thresholds collapse to two:
\[
\Gamma_{\sf{IM}} = \Gamma_{\sf{NZ}_1} = T\underline{q}^r,
\qquad
\Gamma_{\sf{EX}} = \Gamma_{\sf{NZ}_2} = T\overline{q}^r.
\]
With $w_t^{h\ast} = 0$ for all $t$, the NZ-interior aggregate-balance
equation~\eqref{eq:muNZ-interior-proof} reduces to
$T\phi^r(\mu/\pi^w) = \bm{1}^\top \bm{g}$, giving the uniform
optimum
$q_t^{r\ast} = \phi^r(\mu^{\sf{NZ}}/\pi^w) = \bm{1}^\top \bm{g}/T$
at every $t$, and hence
$w_t^{r\ast} = f^r(\bm{1}^\top \bm{g}/T)$ as in~\eqref{eq:WrOnly}. Note
that the strict concavity of $f^r$ forces the per-period RODP setpoints
to coincide; the per-period feasible choice $q_t^r = g_t$ satisfies the
aggregate constraint when $g_t \in [\underline{q}^r, \overline{q}^r]$
but is generally sub-optimal when $f^r$ is strictly concave. In the
linear case $f^r(q) = \alpha^r q$, $f^{\prime r} \equiv \alpha^r$ and any
per-period allocation summing to $\bm{1}^\top \bm{g}$ within the bounds
is equally optimal.

\paragraph{TDP-only case.}
All RODP terms vanish ($w_t^r = 0$ for all $t$, equivalently
$\underline{q}^r = \overline{q}^r = 0$). From~\eqref{eq:GammaDefs_new},
$\Gamma_\sigma = -\sum_t \widetilde{w}_t^h(\delta_\sigma)/\eta_t^h \leq 0$
for every $\sigma \in \{\sf{IM}, \sf{NZ}_1, \sf{NZ}_2, \sf{EX}\}$
(since $\widetilde{w}_t^h(\cdot) \geq 0$). Hence
$\bm{1}^\top \bm{g} \geq 0 \geq \Gamma_{\sf{EX}}$, so the WDP always
operates in the EX regime, and $w_t^{h\ast} = \widetilde{w}_t^h(\pi^-)$
for all $t \in \Tc$, matching Corollary~\ref{corol:RODPTDPonly}.
\hfill\qedsymbol

\subsection{Proof of Theorem~\ref{thm:maxprofit}}
Define the maximized profit under the optimal dispatch in
Theorem~\ref{thm:optimal} as
\[
\Pi^\ast(\bm{g})
:= \Pi\big(\bm{w}^{h\ast}(\bm{g}), \bm{w}^{r\ast}(\bm{g}),
\bm{q}^{h\ast}(\bm{g}), \bm{q}^{r\ast}(\bm{g}), \bm{p}^{h\ast}(\bm{g}); \bm{g}\big),
\]
with $\bm{q}^{h\ast} = \bm{w}^{h\ast} \oslash \bm{\eta}^h$,
$q_t^{r\ast} = (f^r)^{-1}(w_t^{r\ast})$,
$p_t^{h\ast} = (f_t^h)^{-1}(w_t^{h\ast})$, and the aggregate net
consumption $z^\ast := \bm{1}^\top(\bm{q}^{r\ast} - \bm{q}^{h\ast} - \bm{g})$.

Since the profit depends on $\bm{g}$ only through the aggregate
$\bm{1}^\top \bm{g}$ in the IM and EX regions and through the duals
$\mu^{\sf{NZ}_1}, \mu^{\sf{NZ}}, \mu^{\sf{NZ}_2}$ in the respective NZ
sub-segments, we examine $d\Pi^\ast / d(\bm{1}^\top \bm{g})$ region by
region.

\vspace{0.5ex}\noindent\textbf{$\bm{1}^\top \bm{g} < \Gamma_{\sf{IM}}$ (IM).}
From Theorem~\ref{thm:optimal}, $w_t^{h\ast} = \widetilde{w}_t^h(\pi^+)$
and $w_t^{r\ast} = \underline{w}^r$ for all $t$, both independent of
$\bm{g}$. The aggregate net consumption is
$z^\ast = \Gamma_{\sf{IM}} - \bm{1}^\top \bm{g} > 0$, so by the envelope
theorem,
\[
\frac{d\Pi^\ast}{d(\bm{1}^\top \bm{g})}
= -\frac{\partial P}{\partial z}\bigg|_{z^\ast > 0} \cdot \frac{dz^\ast}{d(\bm{1}^\top \bm{g})}
= -\pi^+ \cdot (-1) = \pi^+.
\]

\vspace{0.5ex}\noindent\textbf{$\bm{1}^\top \bm{g} \in [\Gamma_{\sf{IM}}, \Gamma_{\sf{NZ}_1})$ (NZ-lower).}
From Theorem~\ref{thm:optimal}, $w_t^{r\ast} = \underline{w}^r$ and
$w_t^{h\ast} = \widetilde{w}_t^h(\mu^{\sf{NZ}_1})$ at each $t$, with
$z^\ast = 0$. By the envelope theorem applied to the parametrized NZ
subproblem with aggregate dual $\mu^{\sf{NZ}_1}$,
\[
\frac{d\Pi^\ast}{d(\bm{1}^\top \bm{g})} = \mu^{\sf{NZ}_1}(\bm{1}^\top \bm{g}),
\]
with $\mu^{\sf{NZ}_1}$ varying continuously in
$[\delta_{\sf{NZ}_1}, \pi^+]$ across this segment (matching the IM side
at $\Gamma_{\sf{IM}}$ and the NZ-interior side at $\Gamma_{\sf{NZ}_1}$,
per~\eqref{eq:muNZ1-implicit}).

\vspace{0.5ex}\noindent\textbf{$\bm{1}^\top \bm{g} \in [\Gamma_{\sf{NZ}_1}, \Gamma_{\sf{NZ}_2}]$ (NZ-interior).}
From Theorem~\ref{thm:optimal}, both
$w_t^{h\ast} = \widetilde{w}_t^h(\mu^{\sf{NZ}})$ and
$w_t^{r\ast} = f^r(\phi^r(\mu^{\sf{NZ}}/\pi^w))$ vary with
$\bm{g}$ through $\mu^{\sf{NZ}}$, with $z^\ast = 0$ throughout. By the
envelope theorem applied to the parametrized NZ subproblem,
\[
\frac{d\Pi^\ast}{d(\bm{1}^\top \bm{g})} = \mu^{\sf{NZ}}(\bm{1}^\top \bm{g}),
\]
which decreases continuously from $\delta_{\sf{NZ}_1}$ to
$\delta_{\sf{NZ}_2}$ as $\bm{1}^\top \bm{g}$ traverses
$[\Gamma_{\sf{NZ}_1}, \Gamma_{\sf{NZ}_2}]$
(per~\eqref{eq:muNZ-interior-proof}). In the linear case
$f^{\prime r}(q) \equiv \alpha^r$, this collapses to the constant
$f^{\prime r} \pi^w$.

\vspace{0.5ex}\noindent\textbf{$\bm{1}^\top \bm{g} \in (\Gamma_{\sf{NZ}_2}, \Gamma_{\sf{EX}}]$ (NZ-upper).}
By symmetry to the NZ-lower segment, with $w_t^{r\ast} = \overline{w}^r$
and $w_t^{h\ast} = \widetilde{w}_t^h(\mu^{\sf{NZ}_2})$, the same envelope
argument yields
\[
\frac{d\Pi^\ast}{d(\bm{1}^\top \bm{g})}
= \mu^{\sf{NZ}_2}(\bm{1}^\top \bm{g})
\in [\pi^-, \delta_{\sf{NZ}_2}],
\]
matching the NZ-interior side at $\Gamma_{\sf{NZ}_2}$ and the EX side
at $\Gamma_{\sf{EX}}$ (per~\eqref{eq:muNZ2-implicit}).

\vspace{0.5ex}\noindent\textbf{$\bm{1}^\top \bm{g} > \Gamma_{\sf{EX}}$ (EX).}
From Theorem~\ref{thm:optimal}, $w_t^{h\ast} = \widetilde{w}_t^h(\pi^-)$
and $w_t^{r\ast} = \overline{w}^r$ for all $t$, both independent of
$\bm{g}$. The aggregate $z^\ast = \Gamma_{\sf{EX}} - \bm{1}^\top \bm{g} < 0$,
so by the envelope theorem,
\[
\frac{d\Pi^\ast}{d(\bm{1}^\top \bm{g})}
= -\frac{\partial P}{\partial z}\bigg|_{z^\ast < 0} \cdot \frac{dz^\ast}{d(\bm{1}^\top \bm{g})}
= -\pi^- \cdot (-1) = \pi^-.
\]

\vspace{0.5ex}\noindent\textbf{Monotonicity in prices and conversion functions.}
By the envelope theorem applied to the parametric
maximization~\eqref{eq:Optimization},
\begin{align*}
\frac{\partial \Pi^\ast}{\partial \pi^+} &= -[z^\ast]^+ \leq 0, \\
\frac{\partial \Pi^\ast}{\partial \pi^-} &= [z^\ast]^- \geq 0, \\
\frac{\partial \Pi^\ast}{\partial \pi^w} &= \bm{1}^\top(\bm{w}^{h\ast} + \bm{w}^{r\ast}) \geq 0.
\end{align*}
For the conversion functions, pointwise dominance
$\widetilde{f}^r(q) \geq f^r(q)$ for all $q \in \mbbR_+$ relaxes the
RODP input--output relationship without tightening any constraint: the
same RODP electricity input yields weakly more water, so the feasible
profit set expands monotonically. Hence $\Pi^\ast$ is non-decreasing in
$f^r$ in this pointwise sense. By an identical argument applied to the
TDP, pointwise dominance $\widetilde{f}_t^h(p) \geq f_t^h(p)$ for all
$p \in \mbbR_+$ at any $t \in \Tc$ relaxes the TDP fuel-to-water
relationship without tightening any constraint: the same fuel input
yields weakly more water (and, via $q_t^h = w_t^h/\eta_t^h$, weakly
more cogenerated electricity), so $\Pi^\ast$ is non-decreasing in each
$f_t^h$ pointwise. Both monotonicities are strict whenever the
corresponding plant operates in the interior of its bounds at $t$.
\hfill\qedsymbol

\subsection{Proof of Proposition \ref{prop:MPC}}\label{subsec:propMPC}
We reduce $\Qc^{\mathsf{MPC}}_k$ to an instance of~\eqref{eq:Optimization}. The constant terms removed from~\eqref{eq:MPC}, which are the collected water revenue $\pi^w\!\sum_{t<k}(w_t^h+w_t^r)$ and the incurred TDP cost $\sum_{t<k}C_t^h(p_t^h)$, do not affect the maximizer. Writing the payment argument as $\sum_{t=k}^T(q_t^r-q_t^h)-s_k^{\mathsf{eff}}$ with the \emph{effective aggregate renewable}
\begin{equation}\label{eq:seff}
s_k^{\mathsf{eff}}:=\widehat{S}_k-\widetilde{Q}_k,
\end{equation}
problem $\Qc^{\mathsf{MPC}}_k$ is identical to~\eqref{eq:Optimization} posed over the reduced horizon $\Tc_k:=\{k,\ldots,T\}$, with renewable aggregate $s_k^{\mathsf{eff}}$ and unchanged prices, conversion functions, and bounds. By the argument of Lemma~\ref{lem:quasiconcavity} applied over $\Tc_k$, the objective is concave on a convex, compact set, so a global maximizer exists. The demand floor stays slack: any feasible remaining dispatch satisfies $\sum_{t=k}^T(w_t^h+w_t^r)\ge(T{-}k{+}1)(\underline{w}^h+\underline{w}^r)\ge\Wc-\widetilde{W}_k$, where $\widetilde{W}_k:=\sum_{t<k}(w_t^h+w_t^r)\ge(k{-}1)(\underline{w}^h+\underline{w}^r)$ is the exercised water and the last inequality uses the sizing condition; hence $\mathbf{1}^\top(\bm{w}^h+\bm{w}^r)\ge\Wc$ carries no multiplier, exactly as in Theorem~\ref{thm:optimal}.

Theorem~\ref{thm:optimal} thus applies verbatim to $\Qc^{\mathsf{MPC}}_k$ under the substitution $(\Tc,s,\Qc)\mapsto(\Tc_k,s_k^{\mathsf{eff}},\Qc_k)$, where $\Qc_k(\mu,q^r):=\sum_{t=k}^T\!\big(q^r-\widetilde{w}_t^h(\mu)/\eta_t^h\big)$. The per-period setpoints are $w_t^{h\star}=\widetilde{w}_t^h(\mu^k)$ and $q_t^{r\star}=[\phi^r(\mu^k/\pi^w)]_{[\underline{q}^r,\overline{q}^r]}$ for $t\ge k$, and the dual in the net-zero regime is pinned by $\Qc_k(\mu^k,q_t^{r\star}(\mu^k))=s_k^{\mathsf{eff}}$. Adding $\widetilde{Q}_k$ to both sides converts this balance into $\widetilde{\Qc}_k(\mu^k,q_t^{r\star}(\mu^k))=\widehat{S}_k$ by~\eqref{eq:Pk}--\eqref{eq:seff}, converts the reduced thresholds into~\eqref{eq:GammaK}, and maps the regime boundaries on $s_k^{\mathsf{eff}}$ to the stated boundaries on $\widehat{S}_k$. Since $\widetilde{\Qc}_k(\cdot,q_t^{r\star}(\cdot))$ is strictly decreasing in $\mu$ (as $\phi^r$ is strictly decreasing and each $\widetilde{w}_t^h$ is non-decreasing in $\mu$), the net-zero root $\mu^k$ is unique.

\end{document}